\documentclass[onecolumn,draftclsnofoot,nofonttune]{IEEEtran}

\usepackage{amsmath}
\usepackage{amssymb}
\usepackage{amsfonts}
\usepackage{amsthm}

\usepackage[textwidth=20mm]{todonotes}
\usepackage[utf8]{inputenc}
\usepackage[T1]{fontenc}
\usepackage[varg]{txfonts}
\usepackage[mathcal]{euscript}

\usepackage{subfigure}
\usepackage{tikz}
\usepackage{xcolor}
\definecolor{halfgray}{gray}{0.55}
\definecolor{OliveGreen}{rgb}{0,.35,0}
\definecolor{webbrown}{rgb}{.6,0,0}
\definecolor{BrightViolet}{rgb}{0.5,0.2,0.8}
\definecolor{Maroon}{cmyk}{0, 0.87, 0.68, 0.32}
\definecolor{RoyalBlue}{cmyk}{1, 0.50, 0, 0.25}
\definecolor{Black}{cmyk}{0, 0, 0, 0}

\usepackage{acronym}
\usepackage{balance}
\usepackage{bbold}
\usepackage{cancel}
\usepackage{latexsym}
\usepackage{paralist}
\usepackage{wasysym}
\usepackage{xspace}

\usepackage[numbers,sort&compress]{natbib}

\newcommand{\C}{\mathbb{C}}
\newcommand{\R}{\mathbb{R}}

\DeclareMathOperator{\bigoh}{\mathcal O}

\DeclareMathOperator{\diag}{diag}

\DeclareMathOperator{\ess}{ess}
\DeclareMathOperator{\ex}{\mathbb{E}}

\newcommand{\grad}{\nabla}

\DeclareMathOperator{\prob}{\mathbb{P}}

\DeclareMathOperator*{\argmax}{\arg\max}

\DeclareMathOperator{\tr}{tr}
\DeclareMathOperator{\var}{Var}

\newcommand{\dd}{\:d}

\newcommand{\from}{\colon}

\newcommand{\mgeq}{\succcurlyeq}

\newcommand{\pd}{\partial}
\newcommand{\simplex}{\Delta}
\newcommand{\wilde}{\widetilde}

\newcommand{\dis}{\displaystyle}
\newcommand{\txs}{\textstyle}

\newcommand{\insum}{\sum\nolimits}
\newcommand{\inprod}{\prod\nolimits}

\usepackage{algorithm2e_extension}
\usepackage{algorithmic, algorithm}
\SetKwBlock{Repeat}{Repeat}{}

\newtheorem{theorem}{Theorem}

\newtheorem*{corollary*}{Corollary}
\newtheorem{lemma}{Lemma}
\newtheorem{proposition}{Proposition}

\theoremstyle{definition}

\newtheorem*{definition*}{Definition}
\newtheorem{assumption}{Assumption}

\theoremstyle{remark}
\newtheorem{remark}{Remark}
\newtheorem*{remark*}{Remark}

\DeclareMathOperator{\nullity}{nullity}

\newcommand{\bA}{\mathbf{A}}

\newcommand{\bH}{\mathbf{H}}
\newcommand{\bI}{\mathbf{I}}

\newcommand{\bM}{\mathbf{M}}

\newcommand{\bP}{\mathbf{P}}
\newcommand{\bQ}{\mathbf{Q}}
\newcommand{\bR}{\mathbf{R}}

\newcommand{\bU}{\mathbf{U}}
\newcommand{\bV}{\mathbf{V}}
\newcommand{\bW}{\mathbf{W}}

\newcommand{\bY}{\mathbf{Y}}
\newcommand{\bZ}{\mathbf{Z}}

\newcommand{\bp}{\mathbf{p}}
\newcommand{\bq}{\mathbf{q}}
\newcommand{\br}{\mathbf{r}}

\newcommand{\bu}{\mathbf{u}}
\newcommand{\bv}{\mathbf{v}}
\newcommand{\bw}{\mathbf{w}}
\newcommand{\bx}{\mathbf{x}}
\newcommand{\by}{\mathbf{y}}
\newcommand{\bz}{\mathbf{z}}

\newcommand{\bXi}{\mathbf{\Xi}}

\newcommand{\testP}{\bP_{0}}
\newcommand{\testQ}{\bQ_{0}}
\newcommand{\testp}{\bp_{0}}
\newcommand{\testq}{\bq_{0}}
\newcommand{\testx}{\bx_{0}}

\newcommand{\primset}{\mathcal{P}}
\newcommand{\secset}{\mathcal{S}}
\newcommand{\users}{\mathcal{Q}}
\newcommand{\powset}{\strat_{0}}
\newcommand{\covset}{\strat_{+}}
\newcommand{\spectron}{\mathcal{D}}
\newcommand{\cvset}{\mathcal{C}}
\newcommand{\intspectron}{\spectron^{\circ}}

\DeclareMathOperator{\reg}{Reg}
\DeclareMathOperator{\eff}{eff}
\newcommand{\temp}{\eta}
\newcommand{\vartemp}{\gamma}
\newcommand{\choice}{Q}
\newcommand{\gibbs}{G}

\newcommand{\pay}{v}
\newcommand{\vecpay}{\bv}
\newcommand{\bpay}{\bV}
\newcommand{\gradpay}{\bM}
\newcommand{\gradmax}{M}

\newcommand{\score}{y}
\newcommand{\vecscore}{\by}
\newcommand{\bscore}{\bY}
\newcommand{\filter}{\mathcal{F}}

\newcommand{\braket}[2]{\left\langle #1 \middle\vert  #2 \right\rangle}

\newcommand{\ceil}[1]{\left\lceil #1 \right\rceil}
\newcommand{\floor}[1]{\left\lfloor #1 \right\rfloor}

\newcommand{\chan}{\mathcal{K}}
\newcommand{\act}{\mathcal{A}}
\newcommand{\rate}{\Phi}
\newcommand{\strat}{\mathcal{X}}

\newcommand{\negspace}{\!\!\!}

\begin{document}


\title{Transmit without Regrets:
Online Optimization in MIMO\textendash OFDM Cognitive Radio Systems}


\author{%
Panayotis Mertikopoulos,%
~\IEEEmembership{Member,~IEEE},
and
E.~Veronica Belmega,%
~\IEEEmembership{Member,~IEEE}
\thanks{
Manuscript received January 5, 2014;
revised May 19, 2014.}
\thanks{%
This research was supported in part by the European Commission in the framework of the FP7 Network of Excellence in Wireless COMmunications NEWCOM\# (contract no. 318306),
by the French National Research Agency projects
NETLEARN (ANR\textendash 13\textendash INFR\textendash 004)
and GAGA (ANR\textendash 13\textendash JS01\textendash 0004\textendash 01),
and by ENSEA, Cergy\textendash Pontoise, France.
Part of this work was presented at the 7th International Conference on Performance Evaluation and Tools (\textsc{ValueTools} 2013), Turin, Italy, Dec.~2013.}
\thanks{P.~Mertikopoulos is with the French National Center for Scientific Research (CNRS) and the Laboratoire d'Informatique de Grenoble, Grenoble, France;
E.~V.~Belmega is with ETIS/ENSEA\textendash Universit\'e de Cergy-Pontoise\textendash CNRS, Cergy-Pontoise, France.}
}

\maketitle

\begin{abstract}
In this paper, we examine cognitive radio systems that evolve dynamically over time due to changing user and environmental conditions.
To combine the advantages of orthogonal frequency division multiplexing (OFDM) and multiple-input, multiple-output (MIMO) technologies, we consider a MIMO\textendash OFDM cognitive radio network where wireless users with multiple antennas communicate over several non-interfering frequency bands.
As the network's primary users (PUs) come and go in the system, the communication environment changes constantly (and, in many cases, randomly).
Accordingly, the network's unlicensed, secondary users (SUs) must adapt their transmit profiles ``on the fly'' in order to maximize their data rate in a rapidly evolving environment over which they have no control.
In this dynamic setting, static solution concepts (such as Nash equilibrium) are no longer relevant, so we focus on dynamic transmit policies that lead to \emph{no regret}:
specifically, we consider policies that perform at least as well as (and typically outperform) even the best fixed transmit profile in hindsight.
Drawing on the method of matrix exponential learning and online mirror descent techniques, we derive a no-regret transmit policy for the system's SUs which relies only on local channel state information (CSI).
Using this method, the system's SUs are able to track their individually evolving optimum transmit profiles remarkably well, even under rapidly (and randomly) changing conditions.
Importantly, the proposed augmented exponential learning (AXL) policy leads to no regret even if the SUs' channel measurements are subject to arbitrarily large observation errors (the imperfect CSI case),
thus ensuring the method's robustness in the presence of uncertainties.
\end{abstract}


\newacro{CR}{cognitive radio}
\newacro{SISO}{single-input and single-output}
\newacro{MIMO}{multiple-input and multiple-output}
\newacro{UCB}{upper confidence bound}
\newacro{MAC}{multiple access channel}
\newacro{MUI}{multi-user interference-plus-noise}
\newacro{PMAC}{parallel multiple access channel}
\newacro{CSI}{channel state information}
\newacro{CSIT}{channel state information at the transmitter}
\newacro{CDMA}{code division multiple access}
\newacro{FDMA}{frequency division multiple access}
\newacro{DSL}{digital subscriber line}
\newacro{SIC}{successive interference cancellation}
\newacro{SUD}{single user decoding}
\newacro{SINR}{signal-to-interference-and-noise ratio}
\newacro{KKT}{Ka\-rush--Kuhn--Tuc\-ker}
\newacro{IWF}{iterative water-filling}
\newacro{SWF}{simultaneous water-filling}
\newacro{PU}{primary user}
\newacro{SU}{secondary user}
\newacro{iid}[i.i.d.]{independent and identically distributed}
\newacro{OFDMA}{orthogonal frequency-division multiple access}
\newacro{XL}{exponential learning}
\newacro{AXL}{augmented exponential learning}
\newacro{FCC}{Federal Communications Commission}
\newacro{NTIA}{National Telecommunications and Information Administration}
\newacro{GAO}{General Accounting Office}
\newacro{QoS}{quality of service}
\newacro{OFDM}{orthogonal frequency division multiplexing}\acused{OFDM}
\newacro{EW}{exponential weight}
\newacro{OGD}{online gradient descent}
\newacro{OMD}{online mirror descent}


\section{Introduction}
\label{sec:intro}

\IEEEPARstart{T}{he} explosive spread of Internet-enabled mobile devices has turned the radio spectrum into a scarce resource which, if not managed properly, may soon be unable to accommodate the soaring demand for wireless broadband and the ever-growing volume of data traffic and cellphone calls.
Exacerbating this issue, studies by the US \ac{FCC} and the \ac{NTIA} have shown that this vital commodity is effectively squandered through underutilization and inefficient use:
only $15\%$ to $85\%$ of the licensed radio spectrum is used on average, leaving ample spectral voids that could be exploited for opportunistic radio access \cite{GAO04,FCC02}.

In view of the above, the emerging paradigm of \ac{CR} has attracted considerable interest as a promising counter to spectrum scarcity \cite{MM99,ZS07,Hay05,GJMS09}.
At its core, this paradigm is simply a two-level hierarchy between communicating users based on spectrum licensing.
On the one hand, the network's \acp{PU} have purchased spectrum rights but allow others to access it provided that their negotiated \ac{QoS} guarantees are not violated;
on the other hand, the network's \acp{SU} are free-riding on the licensed part of the spectrum, but they have no \ac{QoS} guarantees and must conform to the constraints imposed by the \acp{PU}.
In this way, by opening up the unfilled ``white spaces'' of the licensed spectrum to opportunistic radio access, the overall utilization of the wireless medium can be greatly increased without compromising the performance guarantees that the network's licensed users have already paid for.

Orthogonally to the above, the seminal prediction that \ac{MIMO} technologies can lead to substantial gains in information throughput \cite{FG98,Tel99} opens up additional ways for overcoming spectrum scarcity.
In particular, by employing multiple antennas, it is possible to exploit spatial degrees of freedom in the transmission and reception of radio signals, the only physical limit being the number of antennas that can be deployed on a portable device.
As a result, the existing wireless medium can accommodate greater volumes of data traffic per Hertz without requiring the reallocation (and subsequent re-regulation) of additional frequency bands.

In this paper, we combine these two approaches and focus on dynamic \ac{MIMO} \acl{CR} systems comprising several wireless users (primary and secondary alike) who communicate over multiple non-in\-ter\-fe\-ring channels.
In this evolving (and unregulated) context, the intended receiver of a message has to cope with unwarranted interference from a large number of transmitters, a factor which severely limits the capacity of the wireless system in question.
As a result, given that the system's \acp{SU} cannot rely on contractual \ac{QoS} guarantees to achieve their desired throughput levels, the maximization of their achievable transmission rates under the operational constraints imposed by the network's \acp{PU} becomes a critical issue.

On that account, and given that the theoretical performance limits of \ac{MIMO} systems still elude us (even in basic network models such as the interference channel), a widespread approach is to treat the interference from other users as additive colored noise and to use the mutual information for Gaussian input and noise as a unilateral performance metric \cite{Tel99}.
However, since users cannot be assumed to have full information on the wireless system as it evolves over time (due to the arrival of new users, fluctuations in the \acp{PU}' demand, etc.), they must optimize their signal characteristics ``on the fly'', based only on locally available information.
Hence, our aim is to derive a dynamic transmit policy that allows the system's \acp{SU} to adapt to changes in the wireless medium and to track their individually optimum transmission profiles using only local (and possibly imperfect) \ac{CSI}.

This setting is fairly general and involves cognitive \acp{SU} with significant control over both spatial (\ac{MIMO}) and spectral (\ac{OFDM}) degrees of freedom.
To the best of our knowledge, only special cases of this problem have been considered in a \ac{CR} setting.
For instance, \cite{ZS11,SP10,WSP11} analyzed the case where there is only one channel and the environment is \emph{static} (i.e. the system's \acp{SU} only react to each other and the \acp{PU}' spectrum utilization is fixed);
in this context, \cite{ZS11} characterized the best spatial covariance profile for the interacting \acp{SU} whereas \cite{SP10,WSP11} described how to reach a Nash equilibrium in the resulting non-cooperative game.
On the other hand,
the authors of \cite{NC05,AMTS11,Li09,GKJ10} proposed different learning schemes for optimal channel selection in \emph{dynamic} environments where the \acp{PU}' evolving behavior cannot be anticipated by the system's \acp{SU},
but only in the case where the \acp{SU} are equipped with a single antenna and cannot split power across subcarriers.

Extending the above considerations, our goal in this paper is to derive an adaptive transmit policy for \ac{SU} rate optimization in dynamically evolving \ac{MIMO}\textendash\ac{OFDM} \acl{CR} networks.
In this online optimization framework, the most widely used performance criterion is that of \emph{regret minimization}, a concept which was first introduced by Hannan \cite{Han57} and which has since given rise to a vigorous literature at the interface of optimization, statistics, game theory, and machine learning \textendash\ see e.g. \cite{CBL06,SS11} for a comprehensive survey.
Specifically, in the language of game theory, the notion of (external) regret compares the agent's cumulative payoff over time to what he would have obtained by constantly playing the same action.
Accordingly, the purpose of regret minimization is to devise learning policies that lead to vanishingly small regret against \emph{any} fixed action and \emph{irrespective} of how the agent's environment evolves over time.

In view of the above, we will focus on \emph{no-regret} policies that perform at least as well as the asymptotically best fixed policy in terms of each user's achievable transmission rate \textendash\ despite the fact that the latter cannot be determined by the \acp{SU} when they have no means to anticipate the \acp{PU}' behavior.
In particular, motivated by the no-regret properties of the \ac{EW} algorithm for problems with discrete action sets \cite{CBL06,Vov90,LW94,ACBG02}, we propose an \ac{AXL} approach that can be applied to the continuous regret minimization problem at hand with minimal information requirements.
A key challenge here is that any learning algorithm must respect the problem's semidefiniteness constraints;
as such, an important component of our \ac{AXL} scheme is the continuous-time technique of \emph{matrix exponential learning} that was recently introduced for ordinary (as opposed to online) rate optimization problems in \ac{MIMO} \acp{MAC} \cite{MBM12} \textendash\ and which is in turn closely related to the online mirror descent approach of \cite{SS11} and the matrix regularization techniques of \cite{KSST12}.

Of course, since the \acp{SU}' optimal transmit profile varies over time, the notions of convergence and/or convergence speed are no longer applicable;
instead, the figure of merit is the rate at which the \acp{SU} attain a no-regret state.
In that respect, \ac{AXL} guarantees a worst-case average regret of $\bigoh(T^{-1/2})$ after $T$ epochs, a bound which is well known to be tight \cite{CBL06,SS11}.
Additionally, \ac{AXL} retains its no-regret properties even if the \acp{SU}' channel measurements are subject to arbitrarily large observation errors (the imperfect \ac{CSI} case), thus providing significant performance improvements over more traditional water-filling methods that are sensitive to perfect \ac{CSI}.
As a result, the system's \acp{SU} are able to track their individually optimum transmit profile as it evolves over time remarkably well, even under rapidly (and randomly) changing conditions.

\subsection*{Paper Outline and Summary of Results}

The breakdown of our paper is as follows:
in Section \ref{sec:model}, we introduce our \ac{MIMO}\textendash\ac{OFDM} \acl{CR} network model and the notion of a no-regret transmission policy in the context of \ac{SU} rate optimization.
In Section \ref{sec:components}, we decompose this online rate optimization problem into two components, and we propose a no-regret algorithm for each one.
Specifically, in Section \ref{sec:PA}, we propose an adaptive power allocation policy for the problem's \ac{OFDM} component, whereas in Section \ref{sec:COV}, we derive a dynamic signal covariance policy for the problem's \ac{MIMO} component based on matrix exponential learning.
These components are fused in Section \ref{sec:learning} where we present our \acf{AXL} method for the general \ac{MIMO}-\ac{OFDM} setting and we show that it leads to no regret (Theorem \ref{thm:AXL}).
Importantly, we also show that the \ac{AXL} algorithm retains its no-regret properties even when the users only have imperfect \ac{CSI} at their disposal (Theorem \ref{thm:AXL-stoch}).
This theoretical analysis is validated and supplemented by numerical simulations in Section \ref{sec:numerics} where we also examine the users' ability to track their individually optimum transmit characteristics.
To facilitate presentation, proofs and technical details have been delegated to a series of appendices at the end of the paper.

\section{System Model}
\label{sec:model}

\subsection{The Network Model}

The cognitive radio system that we will focus on consists of a set of non-cooperative wireless \ac{MIMO} users (primary and secondary alike) that communicate over several non-interfering subcarriers by means of an \ac{OFDM} scheme \cite{BGP02,LZ06}.
Specifically, let $\users = \primset \cup \secset$ denote the set of the system's users with $\primset$ (resp. $\secset$) representing the system's primary (resp. secondary) users;
assume further that each user $q\in\users$ is e\-qu\-ip\-ped with $m_{q}$ transmit antennas and that the radio spectrum is partitioned into a set $\chan = \{1,\dotsc,K\}$ of $K$ orthogonal frequency bands \cite{BGP02}.
Then, the aggregate signal $\by_{k}^{s}\in\C^{n_{s}}$ on the $k$-th subcarrier at the intended receiver of the \acl{SU} $s\in\secset$ (assumed equipped with $n_{s}$ receive antennas) will be:
\begin{equation}
\label{eq:signal-multi}
\by_{k}^{s}
	= \bH_{k}^{ss} \bx_{k}^{s}
	+ \insum_{p\in \primset} \bH_{k}^{ps} \bx_{k}^{p}
	+ \insum_{r \in \secset,r\neq s} \bH_{k}^{rs} \bx_{k}^{r}
	+ \bz_{k}^{s},
\end{equation}
where $\bx_{k}^{q}\in\C^{m_{q}}$ is the transmitted message of user $q\in\users$ (primary or secondary) over the $k$-th subcarrier,
$\bH_{k}^{qs}$ is the channel matrix between the $q$-th transmitter and the intended receiver of user $s$,
and $\bz_{k}^{s}\in\C^{n_{s}}$ is the noise in the channel, including thermal, atmospheric and other peripheral interference effects
(and modeled as a non-singular, zero-mean Gaussian vector).
Accordingly, if we focus for simplicity on a specific \ac{SU} and drop the user index $s\in\secset$ in \eqref{eq:signal-multi}, we obtain the signal model
\begin{equation}
\label{eq:signal}
\by_{k} = \bH_{k} \bx_{k} + \bw_{k},
\end{equation}
where $\bw_{k}$ denotes the \ac{MUI} over subcarrier $k\in\chan$ at the intended receiver.

The covariance of $\bw_{k}$ in \eqref{eq:signal} obviously changes over time due to fading, modulations in the \acp{PU}' behavior, etc.;
as a result, employing sophisticated \ac{SIC} techniques at the receiver is highly nontrivial, especially with regards to the system's unregulated \aclp{SU};
Instead, we will work in the \ac{SUD} regime where interference by other users (primary and secondary alike) is treated as additive, colored noise.
In this context, the transmission rate of a user under the signal model \eqref{eq:signal} is given by the familiar expression \cite{Tel99,BGP02}:
\begin{equation}
\label{eq:rate-static}
\txs
\rate(\bP)
	=\insum_{k}
	\big[
	\log\det\big( \bW_{k} + \bH_{k} \bP_{k} \bH_{k}^{\dag} \big) - \log\det\bW_{k}
	\big],
\end{equation}
where:
\begin{enumerate}
\item
$\bW_{k} = \ex\big[\bw_{k} \bw_{k}^{\dag}\big]$ is the multi-user interference-plus-noise covariance matrix over subcarrier $k$.
\item
$\bP_{k} = \ex[\bx_{k} \bx_{k}^{\dag}]$ is the covariance matrix of the user's transmitted signal on subcarrier $k$ and $\bP = \diag(\bP_{1},\dotsc,\bP_{K})$ denotes the user's transmit profile over all subcarriers.
In particular, we will write for convenience:
\begin{equation}
\label{eq:covariance}
\bP_{k} = p_{k} \bQ_{k},
\end{equation}
where $p_{k} = \ex[\bx_{k}^{\dag}\bx_{k}]$ denotes the user's \emph{transmit power} over subcarrier $k$ and $\bQ_{k} = \ex\big[\bx_{k} \bx_{k}^{\dag}\big] \big/  \ex\big[\bx_{k}^{\dag} \bx_{k}\big]$ is his \emph{normalized} signal covariance matrix.
\end{enumerate}

Hence, given that $\bW_k$ may change over time due to evolving user conditions,
we obtain the \emph{time-dependent} objective:
\begin{equation}
\label{eq:rate}
\txs
\rate(\bP; t)
	= \insum_{k} \log\det\big[\bI + \wilde\bH_{k}(t) \, \bP_{k} \, \wilde\bH_{k}^{\dag}(t)\big],
\end{equation}
where the \emph{effective channel matrices} $\wilde \bH_{k}$ are given by 
\begin{equation}
\label{eq:Heff}
\wilde\bH_{k}(t) = \bW_{k}(t)^{-1/2} \, \bH_{k}(t),
\end{equation}
and the time variable $t=1,2,\dotsc$ is assumed discrete (for instance, corresponding to the epochs of a time-slotted system).

Obviously, since we are putting no constraints on the behavior of the system's users,
the evolution of the effective channel matrices $\wilde\bH_{k}(t)$ over time can be quite arbitrary as well.
Formally, we only make the following (minimal) assumptions:
\begin{enumerate}
[{A}1)]
\item
The effective channel matrices $\wilde\bH_{k}(t)$ are bounded for all $t$.

\item
The matrices $\wilde\bH_{k}(t)$ change sufficiently slowly relative to the coherence time of the channel so that the standard results of information theory \cite{Tel99} continue to hold.

\item
\acp{SU} can obtain possibly imperfect (but otherwise unbiased) estimates for $\wilde\bH_{k}$, e.g. by measuring $\bH_{k}$ and probing the intended receiver for the \ac{MUI} covariance matrix $\bW_{k}$.
\end{enumerate}

%

In light of the above, and motivated by the ``white-space filling'' paradigm advocated (e.g. by the \ac{FCC}) as a means to minimize interference by unlicensed users
\cite{FCC02,GAO04,HH10,SP10,SCLH+09}, we will consider the following constraints for the system's \acp{SU}:
\begin{subequations}
\label{eq:constraints}
\begin{enumerate}
[{C}1)]

\item
Bounded total transmit power:
\begin{equation}
\label{eq:constraints-tot}
\txs
\tr(\bP)
	= \insum_{k} p_{k}
	\leq P.
\end{equation}

\item
Constrained transmit power per subcarrier:
\begin{equation}
\label{eq:constraints-chan}
\tr(\bP_{k})
	= p_{k}
	\leq P_{k}.
\end{equation}

\item
Null-shaping constraints:
\begin{equation}
\label{eq:constraints-null}
\bU_{k}^{\dag} \bP_{k} = 0,
\end{equation}
for some tall complex matrix $\bU_{k}$ with full column rank.
\end{enumerate}
\end{subequations}

Of the constraints above, \eqref{eq:constraints-tot} is a physical constraint on the user's total transmit power, \eqref{eq:constraints-chan} imposes a limit on the interference level that can be tolerated on a given subcarrier,
and \eqref{eq:constraints-null} is a ``hard'', spatial version of \eqref{eq:constraints-chan} which guarantees that certain spatial dimensions per subcarrier
are only open to licensed, primary users.
In more detail, \eqref{eq:constraints-chan} is equivalent to limiting the maximal average interference that \acp{SU} are allowed to incur on the primary transmission
while the matrices $\bU_{k}$ of \eqref{eq:constraints-null} are imposed by the \acp{PU} and their columns represent the spatial directions which are forbidden to \ac{SU} transmission.
Such constraints are well-documented in the literature and simply reflect the fact that some carriers or spatial directions per carrier are preferred by the \acp{PU}, so stricter constraints are imposed to limit interference by \acp{SU} (for a more detailed discussion, see e.g. \cite{LZ06,SP10,WSP11} and references therein).

Of course, to maximize \eqref{eq:rate} in the absence of energy awareness considerations, the user must saturate his total power constraint \eqref{eq:constraints-tot} by transmitting at the highest possible (total) power.%
\footnote{Our analysis can be extended to energy-aware objectives where \eqref{eq:constraints-tot} is not saturated, but we will not pursue such directions due to space limitations.}
Thus, the set of admissible transmit profiles for the rate function \eqref{eq:rate} may be expressed as:
\begin{multline}
\label{eq:space}
\txs
\strat
	= \big\{
	\diag(\bP_{1},\dotsc,\bP_{K}):\;
	\bP_{k} \in \C^{m_{k}\times m_{k}},
	\\
	\txs
	\bP_{k} \mgeq 0,
	0 \leq \tr(\bP_{k}) \leq P_{k}
	\text{ and }
	\insum_{k} \tr(\bP_{k}) = P
	\big\},
\end{multline}
where $m_{k} \equiv \nullity(\bU_{k})$ is the number of spatial dimensions that are open to \acp{SU} on subcarrier $k$.
Accordingly, writing $\bP_{k}$ in the decoupled form $\bP_{k} = p_{k} \bQ_{k}$ as in \eqref{eq:covariance}, we obtain the decomposition $\strat = \powset\times\prod_{k} \spectron_{k}$ where
\begin{equation}
\label{eq:powset}
\txs
\powset = \left\{\bp\in\R^{K}:\; 0 \leq p_{k} \leq P_{k},\, \insum_{k} p_{k} = P\right\}
\end{equation}
denotes the set of admissible \emph{power allocation vectors} and
\begin{equation}
\label{eq:spectron}
\spectron_{k} = \big\{\bQ_{k} \in \C^{m_{k}\times m_{k}}:\; \bQ_{k} \mgeq 0, \tr(\bQ_{k}) = 1\big\}
\end{equation}
is the set of admissible \emph{normalized covariance matrices} for subcarrier $k$.
We thus obtain the \emph{online rate maximization problem}:
\begin{equation}
\label{eq:ORM}
\tag{\textup{ORM}}
\begin{aligned}
\text{maximize}
	\quad
	&\txs \rate(\bP; t)\\
\text{subject to}
	\quad
	&\begin{cases}
	\bP = \diag(p_{1} \bQ_{1},\dotsc,p_{K}\bQ_{K}),
	\\
	(p_{1},\dotsc, p_{K}) \in \powset,\;
	\bQ_{k} \in \spectron_{k}.
	\end{cases}
\end{aligned}
\end{equation}

\begin{remark}
In the following sections, we will need the derivatives of $\rate$;
to that end, some matrix calculus yields
\begin{equation}
\label{eq:gradpay}
\frac{\pd\rate}{\pd\bP_{k}^{\ast}}
	\equiv \gradpay_{k}(t)
	= \wilde\bH_{k}^{\dag}(t) \big[ \bI + \wilde\bH_{k}(t) \bP_{k} \wilde\bH_{k}^{\dag}(t)\big]^{-1} \wilde\bH_{k}(t),
\end{equation}
where $\bP_{k}^{\ast}$ denotes the complex conjugate of $\bP_{k}$.
Since the effective channel matrices $\wilde\bH_{k}(t)$ are assumed bounded for all $t$,
the above shows that there exists some $\gradmax>0$ such that:
\begin{equation}
\label{eq:paymax}
\|\gradpay_{k}(t)\|
	\leq \gradmax
	\quad
	\text{for all $k\in\chan$, $\bP\in\strat$, and for all $t\geq0$.}
\end{equation}
\end{remark}

\subsection{Online Optimization and Regret Minimization}

In our setting, there is no direct causal link between the \acp{PU}' behavior and the choices of the \acp{SU}, so the rate function $\rate$
may change arbitrarily over time.
This leads to a ``game against nature'' which is played out as follows:

\begin{enumerate}
\setlength{\itemsep}{1pt}
\setlength{\parskip}{1pt}

\item
At each time slot $t=1,2\dotsc$, the \emph{agent} (i.e. the focal \ac{SU}) selects an \emph{action} (transmit profile) $\bP(t) \in \strat$.

\item
The agent's \emph{payoff} (transmission rate) $\rate(\bP(t);t)$ is determined by nature and/or the behavior of other users (via the effective channel matrices $\wilde\bH_{k}$).

\item
The agent employs some \emph{decision rule} (dynamic transmit policy) to pick a new transmit profile $\bP(t+1) \in \strat$ at stage $t+1$, and the process is repeated until transmission ends.
\end{enumerate}

In this dynamic setting,
static solution concepts are no longer applicable, so the most widely used optimization criterion is that of \emph{regret minimization}, a long-term solution concept which was first introduced by Hannan \cite{Han57} and which has since given rise to an extremely active field of research at the interface of optimization, statistics and theoretical computer science \textendash\ see e.g. \cite{CBL06,SS11} for a survey.
Roughly speaking, the regret compares the payoff obtained by an agent that follows a dynamic policy to the payoff that he would have obtained by constantly choosing the same action over the entire transmission horizon.
More precisely, the \emph{cumulative regret} of the dynamic policy $\bP(t)\in\strat$ with respect to $\testP\in\strat$ is defined as:
\begin{equation}
\label{eq:regret}
\reg_{T}(\testP)
	= \insum_{t=1}^{T} \big[ \rate(\testP;t) - \rate(\bP(t);t) \big],
\end{equation}
i.e. $\reg_{T}(\testP)$ measures the cumulative transmission rate difference up to stage $T$ between a benchmark transmit profile $\testP\in\strat$ and the dynamic policy $\bP(t)$.
The user's \emph{average regret} then is $T^{-1} \reg_{T}(\testP)$ and the goal of regret minimization is to devise a dynamic policy $\bP(t)$ that leads to \emph{no regret}, viz.
\begin{equation}
\label{eq:no-reg}
\limsup\limits_{T\to\infty} \frac{1}{T} \reg_{T}(\testP) \leq 0,
\end{equation}
for all $\testP\in\strat$ and irrespective of the evolution of the objective $\rate(\cdot;t)$ over time.
In other words, if we interpret $\lim_{T\to\infty} T^{-1} \sum_{t=1}^{T} \rate(\testP;t)$ as the long-term average transmission rate of $\testP$, \eqref{eq:no-reg} means that the average data rate of the dynamic transmit policy $\bP(t)$ must be at least as good as that of \emph{any} benchmark profile $\testP\in\strat$.

\begin{remark}
Obviously, if the optimum transmit policy which maximizes \eqref{eq:ORM} could be predicted at every stage $t=1,2,\dotsc$ in an oracle-like fashion, we would have $\reg_{T}(\testP)\leq0$ in \eqref{eq:regret} for all $\testP\in\strat$.
Therfore, the requirement \eqref{eq:no-reg} is fundamental in the context of online optimization because negative regret is a key indicator of tracking the maximum of \eqref{eq:ORM} as it evolves over time.
\end{remark}

\begin{remark}
In the machine learning literature, there exist other notions of regret (such as internal, swap or adaptive regret \cite{HazSes09}) for studying online optimization problems in changing environment.
Due to space limitations, we will focus our theoretical analysis on the external regret formulation \eqref{eq:regret} and we will rely on the numerical simulations of Section \ref{sec:numerics} to show how well our proposed dynamic policies track the evolving maximum of the rate maximization problem \eqref{eq:ORM}.
\end{remark}

\begin{remark}
If the channel matrices are drawn at each realization from an \emph{isotropic} distribution \cite{PCL03}, spreading power uniformly across carriers and antennas is the optimal choice when nature (including the network's \acp{PU}) is actively choosing the worst possible channel realization for the transmitter \cite{PCL03}.
A no-regret policy extends this ``min-max'' concept by ensuring that
the policy's achieved transmission rate is asymptotically as good as that of any fixed transmit profile, including obviously the uniform one as a special case where nature is actively playing against the transmitter \textendash\ e.g. jamming.
\end{remark}

\section{Power Allocation and Signal Covariance Optimization}
\label{sec:components}

To build intuition step-by-step, we will break up the online rate maximization problem \eqref{eq:ORM} in simpler components and we will derive a no-regret transmit policy for each one based on an exponential learning principle.
These policies will then be fused into an adaptive transmit policy for the full \ac{MIMO}\textendash\ac{OFDM} problem in Section \ref{sec:learning}.

\subsection{The \ac{OFDM} Component: Online Power Allocation}
\label{sec:PA}

\subsubsection{A gentle start \textendash\ the case $P_{k}\geq P$}
\label{sec:lowP}

For illustration purposes, we first examine the case where the power-per-channel constraints \eqref{eq:constraints-chan} can be absorbed in the total power constraint \eqref{eq:constraints-tot}, i.e. $P_{k} \geq P$ for all $k\in\chan$.
Also, for scaling purposes, it will be more convenient to consider the normalized power variables
\begin{equation}
\label{eq:relpower}
q_{k} = p_{k}/P.
\end{equation}
With this in mind, if the normalized signal covariance profile $\bQ = \diag(\bQ_{1},\dotsc,\bQ_{K})$ of the focal \ac{SU} is kept fixed, we obtain the online power allocation problem:
\begin{equation}
\label{eq:OPA}
\tag{OPA}
\begin{aligned}
\text{maximize}
	\quad
	&\txs \rate(\bq; t),\\
\text{subject to}
	\quad
	&\bq\in\simplex
\end{aligned}
\end{equation}
where $\simplex = \big\{\bq \in \R_{+}^{K}: \sum_{k=1}^{K} q_{k} = 1 \big\}$ denotes the set of feasible (normalized) power allocation profiles and we write $\rate(\bq;t)$ to highlight the dependence of the rate function \eqref{eq:rate} on the normalized power allocation profile $\bq\in\simplex$ (instead of $\bP\in\strat$).

A special case of this problem is when the user cannot split power across subcarriers and can only choose one channel on which to transmit.
Essentially, this channel selection framework boils down to the famous ``multi-armed bandit'' problem of \cite{Rob52} (see e.g. \cite{CBL06,SS11} for a review).
As a result, much recent work on \ac{CR} networks \cite{AMTS11,Li09,GKJ10} has been focused on no-regret channel selection algorithms based on $Q$-learning \cite{Li09} or \ac{UCB} techniques \cite{AMTS11}.

Unfortunately, these techniques are inherently discrete in nature, so it is not clear how to extend them to the continuous context of \eqref{eq:OPA}.
Instead, motivated by the \acl{EW} algorithm introduced in \cite{Vov90,LW94,ACBG02} for sequence prediction, our approach consists of scoring each channel over time and then allocating power proportionally to the exponential of these scores.
In particular, inspired by the analysis of \cite{MBML12}, each channel will be scored by means of the \emph{marginal utilities}:
\begin{equation}
\label{eq:pay-PA}
\pay_{k}
	= \frac{\pd\rate}{\pd q_{k}}
	= P \frac{\pd\rate}{\pd p_{k}}
	= P \cdot \tr\big[\gradpay_{k} \bQ_{k}\big],
\end{equation}
where $\bQ_{k}\in\spectron_{k}$ is the user's (fixed) covariance matrix and $\gradpay_{k}$ is given by \eqref{eq:gradpay}.
We thus obtain the \acl{XL} power allocation policy:
\begin{equation}
\label{eq:XLPA}
\tag{XL-PA}
\begin{aligned}
\score_{k}(t)
	&= \score_{k}(t-1) + \pay_{k}(t),
	\\
q_{k}(t+1)
	&= \frac{\exp\big(\temp t^{-1/2} \score_{k}(t)\big)}{\sum_{\ell} \exp\big(\temp t^{-1/2} \score_{\ell}(t)\big)},
\end{aligned}
\end{equation}
where $\temp>0$ is a learning rate parameter and
the $\sqrt{t}$ factor has been included to moderate very sharp score differences.

Our first result is that \eqref{eq:XLPA} performs asymptotically as well as \emph{any} fixed power allocation profile $\testq\in\simplex$:

\begin{proposition}
\label{prop:XLPA}
If $P_{k} \geq P$ for all $k\in\chan$, the policy \eqref{eq:XLPA} leads to no regret.
Specifically, for every $\testq\in\simplex$,
and independently of the system's evolution over time,
we have
\begin{equation}
\label{eq:reg-XLPA}
\frac{1}{T}\reg_{T}(\testq)
	\leq \frac{1}{\sqrt{T}} \left(\frac{\log K}{\temp} + 4 P^{2}\gradmax^{2} \temp \right),
\end{equation}
with $\gradmax$ given by \eqref{eq:paymax}.
\end{proposition}

\begin{IEEEproof}
See Appendices \ref{app:PA} and \ref{app:discrete}.
\end{IEEEproof}

\setcounter{remark}{0}


\begin{remark}
The use of the marginal utilities \eqref{eq:pay-PA} in the exponential learning policy \eqref{eq:XLPA} can be compared to the online gradient descent algorithm introduced in \cite{Zin03} where the learner tracks the gradient of his evolving objective and projects back to the problem's feasible set when needed.
We did not take such an approach because projections are numerically unstable \cite{Can00} and can become quite costly from a computational standpoint (the problem's constraints would have to be checked individually at every iteration).
Nonetheless, the exponential approach of \eqref{eq:XLPA} has strong ties to the method of online \emph{mirror} descent \cite{SS11} which we discuss later.
\end{remark}

\subsubsection{The general case}
\label{sec:highP}

The dynamic power allocation policy \eqref{eq:XLPA} concerns the case where the power-per-channel constraints \eqref{eq:constraints-chan} can be absorbed in the total power constraint \eqref{eq:constraints-tot}.
Otherwise, if $P_{k}<P$ for some channel $k\in\chan$ (e.g. if certain \acp{PU} have very low interference tolerance on their licensed channels), \eqref{eq:XLPA} cannot be employed ``as is'' because it does not respect the constraint $p_{k} \leq P_{k}$.
When this is the case, the analysis of Appendix \ref{app:PA-mod} yields the modified policy:
\begin{equation}
\label{eq:XLPA-mod}
\tag{\ref{eq:XLPA}$'$}
\begin{aligned}
\score_{k}(t)
	&= \score_{k}(t-1) + \pay_{k}(t),
	\\
p_{k}(t+1)
	&= P_{k} \left(1 + \exp(\lambda - \temp t^{-1/2}\score_{k})\right)^{-1}
\end{aligned}
\end{equation}
where $\lambda>0$ is defined implicitly so that \eqref{eq:constraints-tot} is satisfied:
\begin{equation}
\label{eq:lambda}
\txs
P = \sum_{k\in\chan} P_{k} \left(1 + \exp(\lambda - \temp t^{-1/2}\score_{k})\right)^{-1}.
\end{equation}

Just like \eqref{eq:XLPA}, \eqref{eq:XLPA-mod} exhibits exponential sensitivity to the scores $\score_{k}$ modulo a normalization factor corresponding to the constraints \eqref{eq:constraints-tot} and \eqref{eq:constraints-chan}.
Since the RHS of \eqref{eq:lambda} is strictly decreasing in $\lambda$, it is then easy to calculate the value of $\lambda$ itself, e.g. by performing a line search for $e^{\lambda}$ \cite{Can00}.%
\footnote{See also \cite{MB13} for a closed-form expression of \eqref{eq:XLPA-mod} based on a modified version of the replicator equation of evolutionary game theory.}
We thus get:

\begin{proposition}
\label{prop:XLPA-mod}
The policy \eqref{eq:XLPA-mod} leads to no regret.
In particular, for every $\testp\in\powset$, the user's regret is bounded by
\begin{equation}
\label{eq:reg-XLPA-mod}
T^{-1}\reg_{T}(\testp)
	\leq \bigoh\big(T^{-1/2}\big),
\end{equation}
irrespective of the system's evolution over time.
\end{proposition}

\begin{IEEEproof}
See Appendix \ref{app:PA-mod}.
\end{IEEEproof}

\setcounter{remark}{0}

\begin{remark*}
We should note here that \eqref{eq:XLPA-mod} is not equivalent to \eqref{eq:XLPA} if $P_{k} \geq P$;
instead, \eqref{eq:XLPA} should be viewed as a simpler alternative to \eqref{eq:XLPA-mod} that can be employed whenever the maximum power-per-channel constraints \eqref{eq:constraints-chan} can be subsumed in the total power constraint \eqref{eq:constraints-tot}.
For convenience,
we will present our results in the simpler case $P_{k} \geq P$ and we will rely on a series of remarks to translate these remarks to the regime $P_{k} < P$ (cf. Appendices \ref{app:PA} and \ref{app:PA-mod}).
\end{remark*}

\subsection{The \ac{MIMO} Component: Signal Covariance Optimization}
\label{sec:COV}

If the user's power allocation profile $\bp=(p_{1},\dotsc,p_{K})$ remains fixed throughout the transmission horizon, \eqref{eq:ORM} boils down to the online signal covariance optimization problem:
\begin{equation}
\label{eq:OCOV}
\tag{OCOV}
\begin{aligned}
\text{maximize}
	\quad
	&\txs \rate(\bQ; t),\\
\text{subject to}
	\quad
	&\txs
	\bQ_{k}\mgeq0,\; \tr(\bQ_{k}) = 1,
\end{aligned}
\end{equation}
where we now use the notation $\rate(\bQ;t)$ to highlight the dependence of the user's transmission rate \eqref{eq:rate} on the normalized covariance matrix $\bQ = \diag(\bQ_{1},\dotsc,\bQ_{K})\in \covset \equiv\prod_{k} \spectron_{k}$.

A key challenge in \eqref{eq:OCOV} is that any learning algorithm must respect the problem's (implicit) semidefiniteness constraints $\bQ_{k} \mgeq 0$.
To that end, motivated by the analysis of \cite{MBM12} (see also the matrix regularization approach of \cite{KSST12}),
we will consider the \emph{matrix exponential learning} policy
\begin{equation}
\label{eq:XLCOV}
\tag{XL-COV}
\begin{aligned}
\bscore_{k}(t)
	&= \bscore_{k}(t-1) + \bpay_{k}(t),
	\\
\bQ_{k}(t+1)
	&= \frac{\exp\big(\temp t^{-1/2} \bscore_{k}(t)\big)}{\tr\big[\exp\big(\temp t^{-1/2} \bscore_{k}(t)\big)\big]},
\end{aligned}
\end{equation}
where the matrix-valued gradient payoff $\bpay_{k}$ is defined as:
\begin{equation}
\label{eq:pay-COV}
\bpay_{k}
	= \frac{\pd\rate}{\pd \bQ_{k}^{\ast}}
	= p_{k} \gradpay_{k},
\end{equation}
and $\gradpay_{k}$ is given by \eqref{eq:gradpay}.
Intuitively, \eqref{eq:XLCOV} reinforces the spatial directions that peform well by increasing the corresponding eigenvalues while the $t^{-1/2}$ factor keeps the eigenvalues of $\bQ_{k}$ from approaching zero too fast \cite{KM14}.
Along these lines, our analysis in Appendix \ref{app:COV} yields:

\begin{proposition}
\label{prop:XLCOV}
The dynamic transmit policy \eqref{eq:XLCOV} leads to no regret in the online signal covariance optimization problem \eqref{eq:OCOV}.
In particular, for every $\testQ\in\covset \equiv \prod_{k} \spectron_{k}$, and irrespective of the system's evolution over time, we have:
\begin{equation}
\label{eq:reg-XLCOV}
\frac{1}{T}\reg_{T}(\testQ)
	\leq \frac{1}{\sqrt{T}} \left(\frac{\sum_{k=1}^{K} \log m_{k}}{\temp} + 4 P^{2} \gradmax^{2} \temp\right),
\end{equation}
where $m_{k}$ is the number of spatial degrees of freedom left open by the constraint \eqref{eq:constraints-null}.
\end{proposition}

\setcounter{remark}{0}



\section{Learning in the Full \ac{MIMO}\textendash\ac{OFDM} Problem}
\label{sec:learning}

\subsection{Augmented Exponential Learning}
\label{sec:AXL}

Based on the analysis of the previous section, we derive here a dynamic no-regret policy for the full \ac{MIMO}\textendash\ac{OFDM} problem \eqref{eq:ORM}.
Working for simplicity with the special case $P_{k}\geq P$, \eqref{eq:XLPA} and \eqref{eq:XLCOV} yield the dynamic transmit policy:

\begin{algorithm}[H]
{\sf
\vspace{2pt}
Parameter:
$\temp>0$.\\
Initialize:
$t \leftarrow 0$;
channel scores $\score_{k} \leftarrow 0$, $\bscore_{k}\leftarrow0$.
\\
\Repeat
{$t \leftarrow t+1$;
\\
\ForEach
{channel $k \in \chan$}
{%
set
\(
\dis
\begin{cases}
p_{k} \leftarrow P \exp\big(\temp t^{-1/2} \score_{k} \big)\big/\sum_{\ell} \exp\big(\temp t^{-1/2} \score_{\ell}\big);
\\
\bQ_{k} \leftarrow \exp\big(\temp t^{-1/2} \bscore_{k}\big) \big/ \tr\big[\exp\big(\temp t^{-1/2} \bscore_{k}\big)\big];
\end{cases}
\)}
\ForEach
{channel $k \in \chan$}
{%
measure $\gradpay_{k} \leftarrow \wilde\bH_{k}^{\dag} \big[ \bI + p_{k} \wilde\bH_{k} \bQ_{k} \wilde\bH_{k}^{\dag}\big]^{-1} \wilde\bH_{k}$;
\\
update scores:
$\dis
\begin{cases}
y_{k} \leftarrow y_{k} + P\tr[\gradpay_{k} \bQ_{k}];
\\
\bscore_{k} \leftarrow \bscore_{k} + p_{k} \gradpay_{k};
\end{cases}$
} 
\vspace{1ex}
until transmission ends.} 
} 
\caption{Augmented Exponential Learning (\acs{AXL})}
\label{algo.AXL}
\end{algorithm}

\setcounter{remark}{0}

The \acf{AXL} algorithm above will be our main focus, so a few remarks are in order:

\begin{remark}
From an implementation point of view, \ac{AXL} has the following desirable properties:
\begin{enumerate}
[(P1)]
\setlength{\itemsep}{1pt}
\setlength{\parskip}{1pt}
\item
It is \emph{distributed}:
each \ac{SU} only needs to update his individual transmit policy using local \ac{CSI} (the matrices $\wilde\bH_{k}$).

\item
It is \emph{asynchronous}:
there is no need for a global update timer to synchronize the system's \acp{SU}.

\item
It is \emph{stateless}:
the \acp{SU} do not need to know the state of the system (e.g. the network's topology), and/or be aware of each other's actions.

\item
It is \emph{reinforcing}:
the \acp{SU} tend to increase their unilateral transmission rates.
\end{enumerate}
\end{remark}

\begin{remark}
If the maximum power-per-channel constraints imposed on the network's \acp{SU} do not satisfy the condition $P_{k} \geq P$ for all $k\in\chan$, the power update step of \ac{AXL} must be modified:
specifically, the exponential allocation rule $p_{k} \leftarrow P \exp(\temp t^{-1/2} \score_{k})\big/\sum_{\ell} \exp(\temp t^{-1/2} \score_{\ell})$ must be replaced by the update rule of \eqref{eq:XLPA-mod}, i.e. by setting $p_{k} \leftarrow P_{k} \big[1 + \exp(\lambda - \temp t^{-1/2} \score_{k})\big]^{-1}$.
To simplify our presentation, we will keep the assumption $P_{k} \geq P$ with the implicit understanding that if $P_{k} < P$ for some $k\in\chan$, then it is the modified version of \ac{AXL} that should be used instead.
\end{remark}

With all this in mind, our main result is that the \ac{AXL} algorithm leads to no regret if $P_{k}\geq P$ for all channels:

\begin{theorem}
\label{thm:AXL}
The adaptive transmit policy generated by \ac{AXL} leads to no regret in the online rate maximization problem \eqref{eq:ORM}.
In particular, for every fixed transmit profile $\testP \in \strat$, and independently of how the system's rate function \eqref{eq:rate} evolves over time, the user's regret is bounded by:
\begin{equation}
\label{eq:reg-RM}
\frac{1}{T}\reg_{T}(\testP)
	\leq \frac{1}{\sqrt{T}} \left(\frac{\log K + \insum_{k=1}^{K} \log m_{k}}{\temp} + 4 P^{2} \gradmax^{2} \temp\right),
\end{equation}
where $\gradmax$ is given by \eqref{eq:paymax} and $m_{k}$ is the number of spatial dimensions that are left open to \acp{SU} by the constraint \eqref{eq:constraints-null}.
\end{theorem}

\begin{IEEEproof}
See Appendices \ref{app:AXL} and \ref{app:discrete}.
\end{IEEEproof}

\setcounter{remark}{0}

\begin{remark}
As we already explained, if $P_{k} < P$ for some $k\in\chan$, the power update step in the \ac{AXL} algorithm should be replaced by the power allocation rule \eqref{eq:XLPA-mod}.
In this case, \ac{AXL} still guarantees an $\bigoh(T^{-1/2})$ regret bound  but the exact expression is more complicated (see Appendix \ref{app:PA-mod} for the details).
\end{remark}

\begin{remark}
The proof of Theorem \ref{thm:AXL} relies on a deep connection between \eqref{eq:XLPA} and \eqref{eq:XLCOV} with the Gibbs\textendash Shannon and von Neumann entropy functions respectively.
In fact, as we shall see in Appendices \ref{app:PA}\textendash\ref{app:PA-mod}, our approach is intimately related to the Hessian\textendash Riemannian optimization method of \cite{ABB04} and the online mirror descent techniques of \cite{SS11,KSST12}.
Unfortunately, a full description of these methods requires the introduction of significant technical apparatus, so we will not discuss them at length;
for a detailed account, the reader is instead referred to \cite{SS11,KM14}.
\end{remark}

\begin{remark}
It should also be noted that the bound \eqref{eq:reg-RM} is not the sum of the bounds \eqref{eq:reg-XLPA} and \eqref{eq:reg-XLCOV}.
As we show in Appendices \ref{app:AXL} and \ref{app:discrete}, the reason for this is that Theorem \ref{thm:AXL} is \emph{not} a corollary of Propositions \ref{prop:XLPA} and \ref{prop:XLCOV} but, rather, a combination of these two independent results.
\end{remark}

\begin{remark}
In practice, the learning parameter $\temp$ of the \ac{AXL} algorithm can be tuned freely by the user.
As such, if the user can estimate ahead of time the quantity $\gradmax$ (which can be seen as an effective bound on the gradient matrices $\gradpay_{k}$ over time), $\temp$ can be chosen so as to optimize the regret guarantee \eqref{eq:reg-RM} \textendash\ thus leading to lower regret levels faster.
Specifically, some calculations along the lines of \cite{KM14} show that the optimal choice of $\temp$ which minimizes the RHS of \eqref{eq:reg-RM} is:
\begin{equation}
\label{eq:eta-opt}
\txs
\temp
	= \frac{1}{2} P\gradmax \left(\log K + \insum_{k} \log m_{k}\right)^{1/2},
\end{equation}
which then leads to the optimized regret guarantee:
\begin{equation}
\label{eq:reg-opt}
\txs
\reg_{T}(\testP)
	\leq 4P \gradmax \big(\log K + \insum_{k} \log m_{k}\big)^{1/2} T^{1/2}.
\end{equation}

This bound resembles the bound derived in \cite{KSST12} for learning processes that stop after a predetermined number of steps;
that being said (and in contrast to Theorem \ref{thm:AXL}), unless some sort of ``doubling correction'' is used \cite{CBL06}, the method proposed in \cite{KSST12} may lead to positive regret in an infinite horizon setting (such as the one we are considering here).
On the other hand, this also shows that if the user can estimate his transmission horizon in advance (instead of having an infinite backlog of data to transmit), then he can use \ac{AXL} with constant parameter $\temp$ given by \eqref{eq:eta-opt} and still enjoy the optimal regret guarantee \eqref{eq:reg-opt}.
\end{remark}

\begin{remark}
Finally, we note that the optimal bound \eqref{eq:reg-opt} is asymptotically tight with respect to $T$ but not necessarily with respect to the dimensionality of the problem.
In particular, the analysis of \cite{CBL06,SS11} shows that the best bound that can be guaranteed against an adversarial nature is $\bigoh(\sqrt{T})$;
furthermore, if the state space of the problem is a simplex of dimension $K$, the tightest possible bound is $\bigoh(\log K)$ \cite{CBL06}.
In this way, the $\log K$ factor of \eqref{eq:reg-opt} is tight;
we conjecture that the same holds for the $\log m_{k}$ factors because the covariance spectrahedrons $\spectron_{k}$ are simply the product of a simplex with dimension $m_{k}$ with the space of unitary matrices.
At any rate, the bound \eqref{eq:reg-opt} only tightens against an adversarial nature, so, in practical situations, we expect the user's regret to decay much more rapidly (cf. the numerical simulations of Section \ref{sec:numerics}).
\end{remark}

\subsection{Learning with Imperfect Channel State Information}
\label{sec:AXL-stoch}

In practice, a major challenge occurs if the user does not have perfect \ac{CSI} with which to calculate the matrix gradients \eqref{eq:gradpay} that are needed to run the \ac{AXL} algorithm.
To wit, since these gradients are determined by the effective channel matrices $\wilde\bH_{k} = \bW_{k}^{-1/2} \bH_{k}$, imperfect measurements of the actual channel matrices $\bH_{k}$ or of the multi-user interference-plus-noise covariance matrices $\bW_{k}$ would invariably interfere with each update cycle.
Accordingly, our aim in this section is to study the robustness of \ac{AXL} in the presence of measurement errors.

To account for as wide a range of errors as possible, we will assume that at each update period $t=1,2,\dotsc$, the user can only observe a noisy estimate
\begin{equation}
\label{eq:gradpay-stoch}
\hat\gradpay_{k}(t) = \gradpay_{k}(t) + \bXi_{k}(t)
\end{equation}
of $\gradpay_{k}(t)$, where the noise process $\bXi_{k}(t)$ represents a random and unbiased observational error (not necessarily \acs{iid}).
Formally:
\begin{assumption}
We assume that the observation error $\bXi_{k}$ is:
\begin{enumerate}
\item
\emph{Bounded}:
$\|\bXi_{k}(t)\| \leq \Sigma$ \textup(a.s.\textup) for some $\Sigma>0$ and for all $t$.
\item
\emph{Unbiased}:
$\ex\big[\bXi_{k}(t) \vert \filter_{t-1}\big] = 0$
where $\filter = \{\filter_{t}\}_{t\geq1}$ denotes the history of the user's choices.
\end{enumerate}
\end{assumption}
Remarkably, as long as there is no systematic bias in the user's measurements, the \ac{AXL} algorithm still leads to no regret, even in the presence of \emph{arbitrarily large} observation errors:

\begin{theorem}
\label{thm:AXL-stoch}
The \ac{AXL} algorithm with noisy observations $\hat\gradpay_{k}$ of the form \eqref{eq:gradpay-stoch} leads to no regret \textup(a.s.\textup).
Specifically, if $\|\bXi_{k}\| \leq \Sigma$, then, for all $\testP \in \strat$ and for all $z>0$:
\vspace{5pt}
\begin{enumerate}
[\noindent\textup(i\textup)]
\addtolength{\itemsep}{3pt}

\item
The user's expected regret is bounded by:
\begin{equation}
\label{eq:reg-stoch-avg}
\ex\left[T^{-1}\reg_{T}(\testP)\right]
	\leq R T^{-1/2}.
\end{equation}

\item
The user's realized regret is bounded by the perfect \ac{CSI} guarantee of \ac{AXL} with exponentially high probability:
\begin{equation}
\label{eq:reg-stoch-largedev}
\prob\left( \frac{1}{T} \reg_{T}(\testP) \leq \frac{R}{\sqrt{T}} + z \right)
	\geq 1 - \exp\left(- \frac{z^{2}T}{2 D^{2} \Sigma^{2}}\right),
\end{equation}
\end{enumerate}
where $D>0$ is a constant and $R$ is the deterministic guarantee \eqref{eq:reg-RM} of \ac{AXL} under perfect \ac{CSI}, viz.:
\begin{equation}
\label{eq:reg-det}
\txs
R
	= \temp^{-1}\cdot\big(\log K + \insum_{k} \log m_{k}\big) + 4 P^{2} \gradmax^{2} \temp.
\end{equation}
\end{theorem}

Theorem \ref{thm:AXL-stoch} (proven in Appendix \ref{app:stochastic}) shows that \ac{AXL} guarantees an $\bigoh(T^{-1/2})$ bound on the user's regret with high probability, even under measurement errors of arbitrarily high magnitude.
Accordingly, a few remarks are in order:

\setcounter{remark}{0}

\begin{remark}
The first- and second-order statistics of the measured gradients $\hat\gradpay_{k}$ play different roles in the presence of imperfect \ac{CSI}:
the expected value $\ex\big[\hat\gradpay_{k}\big] = \gradpay_{k}$ of $\hat\gradpay_{k}$ controls the expected regret guarantee of \ac{AXL} via \eqref{eq:reg-stoch-avg}, whereas the variance $\var\big(\hat\gradpay_{k}\big) = \ex\big[\|\bXi_{k}\|^{2}\big]$ of $\hat\gradpay_{k}$ controls the deviations of the regret from its ``bulk'' behavior \textendash\ but has no impact on the expected regret of \ac{AXL}.
\end{remark}

\begin{remark}
Note also that Theorem \ref{thm:AXL} is recovered by \eqref{eq:reg-stoch-largedev} in the deterministic limit $\Sigma\to0^{+}$:
the probability that the user's regret exceeds the determinstic guarantee $R/\sqrt{T}$ converges uniformly to $0$ as $\Sigma\to0^{+}$.
\end{remark}


\section{Numerical Results}
\label{sec:numerics}

%

To validate the predictions of Section \ref{sec:learning} for the \ac{AXL} algorithm, we conducted extensive numerical simulations from which we illustrate here a selection of the most representative scenarios \textendash\ though the observations made below remain valid in most typical mobile wireless environments.

\begin{figure*}[htbp]
\centering
\subfigure
[No regret under \acl{AXL}.]
{\label{fig:regret-AXL}
\includegraphics[width=.49\textwidth]{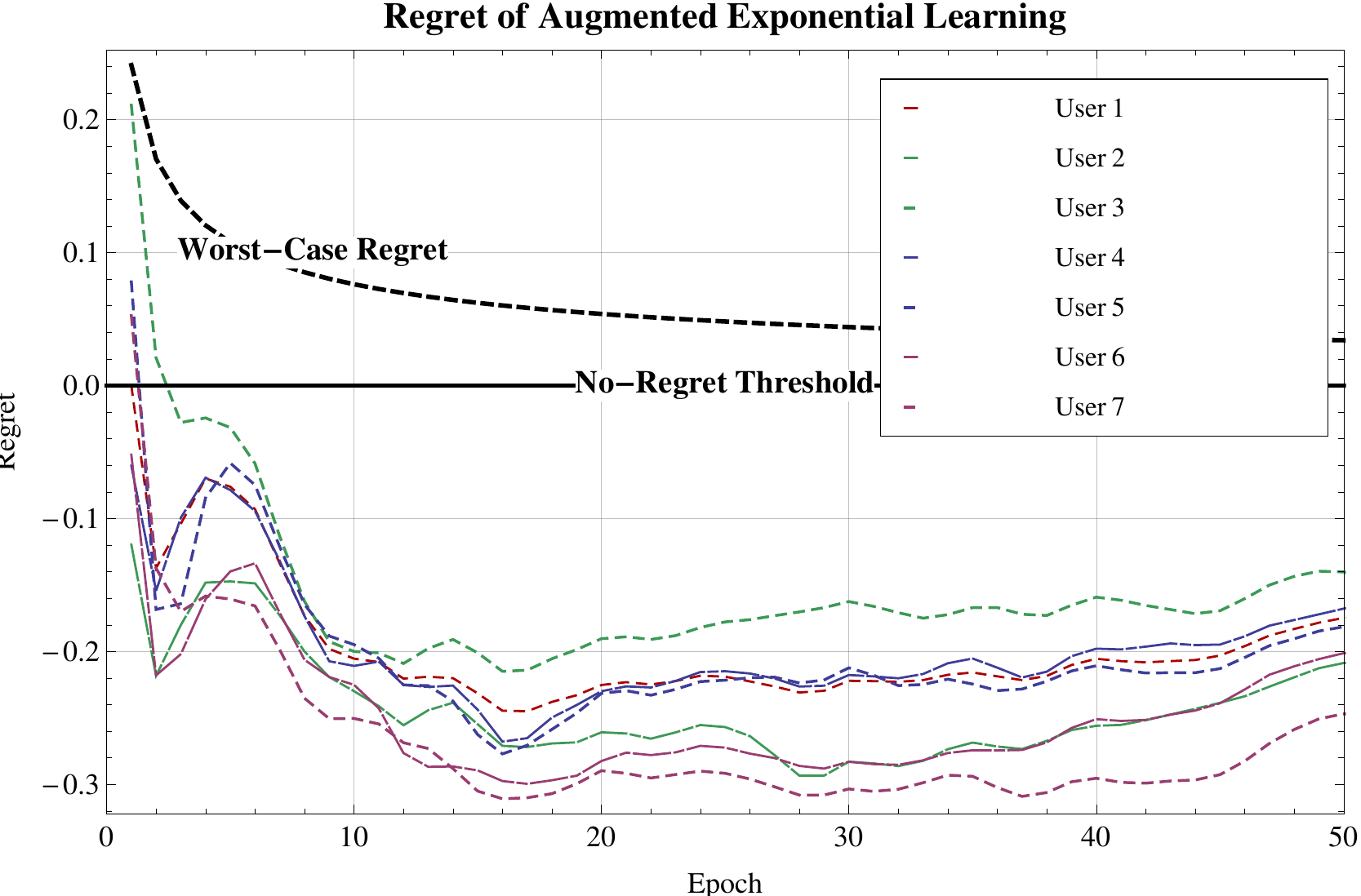}}
\hfill
\subfigure
[Positive regret under randomized power allocation.]
{\label{fig:regret-rand}
\includegraphics[width=.48\textwidth]{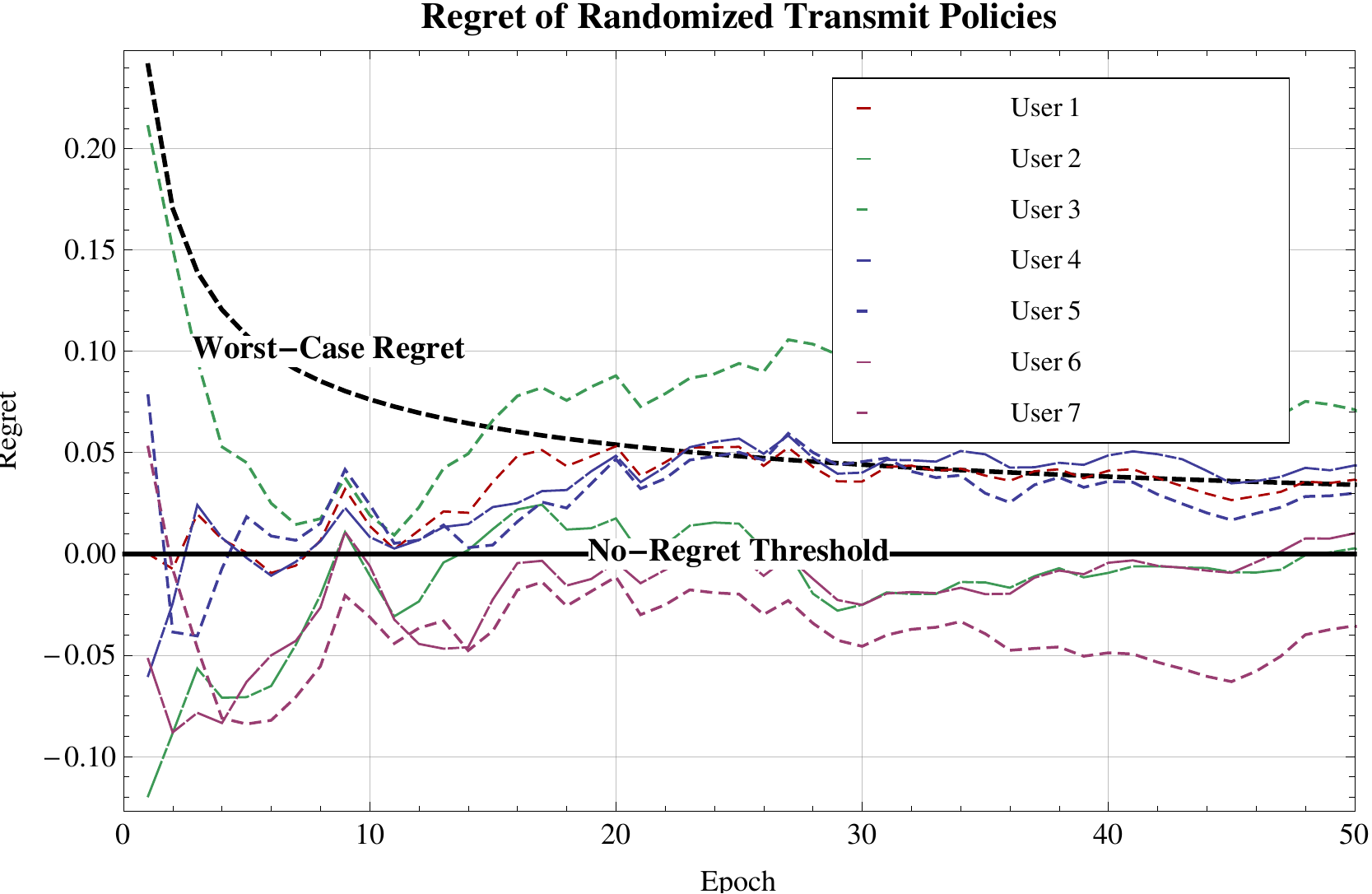}}
\caption{The long-term regret induced by \acl{AXL} and a random sampling transmit policy (Figs \ref{fig:regret-AXL} and \ref{fig:regret-rand} respectively) for different users (see text for details).
In tune with Theorem \ref{thm:AXL}, \ac{AXL} quickly falls below the no-regret threshold whereas the randomized policy \eqref{eq:random} leads to positive regret for several users (in both figures the dashed ``worst-case regret'' curve represents the regret guarantee \eqref{eq:reg-RM} of the \ac{AXL} algorithm).}
\label{fig:regret}
\end{figure*}

In Fig.~\ref{fig:regret}, we simulated a network consisting of $10$ \acp{PU} and $40$ \acp{SU}, all equipped with $m_{k} = 3$ transmit/receive antennas, and communicating over $K=256$ orthogonal subcarriers with a base frequency of $\nu = 2\,\mathrm{GHz}$.
Both the \acp{PU} and the \acp{SU} were assumed to be mobile with a speed between $3$ and $5$ km/h (pedestrian movement), and the channel matrices $\bH_{k}^{qs}$ of \eqref{eq:signal} were modeled after the well-known Jakes model for Rayleigh fading \cite{CCGH+07}.
For simplicity, we assumed that the \acp{PU} were going online and offline following a Poisson process (representing exponential arrivals with exponential call times), while the simulated \acp{SU} employed the \ac{AXL} algorithm with $\temp=1$ and an update epoch of $\delta = 5\,\textrm{ms}$.%
\footnote{We did not optimize the choice of $\temp$ because we wanted to focus on the case where the network's \acp{SU} have minimal information.}
We then calculated the maximum regret induced by the \ac{AXL} for every \ac{SU} with respect to the uniform transmit profile (where power is spread equally across antennas and frequency bands) and all possible combinations of spreading power uniformly across subcarriers while keeping one or two transmit dimensions closed (we plotted the regret for only $7$ \acp{SU} in order to reduce graphical clutter).
The results of these simulations were plotted in Fig.~\ref{fig:regret-AXL}:
as predicted by Theorem \ref{thm:AXL}, \ac{AXL} leads to no regret and falls below the no-regret threshold within a few epochs, indicating that its average performance is strictly better than any of the benchmark transmit profiles.

For comparison purposes, we also simulated the same scenario but with the \acp{SU} employing a randomized transmit policy.
In particular, motivated by \cite{PCL03}, we simulated the randomized transmit policy:
\begin{equation}
\label{eq:random}
\begin{aligned}
\bQ_{k}(t+1)
	&= (1-r) \bQ_{k}(t) + r \bR_{k}(t),
	\\
\bQ_{k}(0)
	&= m_{k}^{-1} \bI,
\end{aligned}
\end{equation}
where the matrix $\bR_{k}(t)$ is drawn uniformly from the spectrahedron $\spectron_{k}$ of $m_{k}\times m_{k}$ positive-definite matrices with unit trace, and $r\in[0,1]$ is a discount parameter interpolating between the uniform distribution $\bQ_{k} \propto \bI$ for $r=0$ and the completely random policy $\bR_{k}$ for $r=1$ (in our simulations, we took $r=0.9$).
Even though this dynamic transmit policy is sampling the state space essentially uniformly for large values of $r$, Fig.~\ref{fig:regret-rand} shows that several \acp{SU} end up having positive regret.
We thus see that the no-regret property of \ac{AXL} is not a spurious artifact of exploring the problem's state space in a uniform way, but it is inextricably tied to the underlying learning mechanism.

\begin{figure*}[htbp]
\centering
\subfigure
[Performance of \ac{AXL} with average user velocity $v=5~\mathrm{km/s}$.]
{\label{fig:tracking-5}
\includegraphics[width=.485\textwidth]{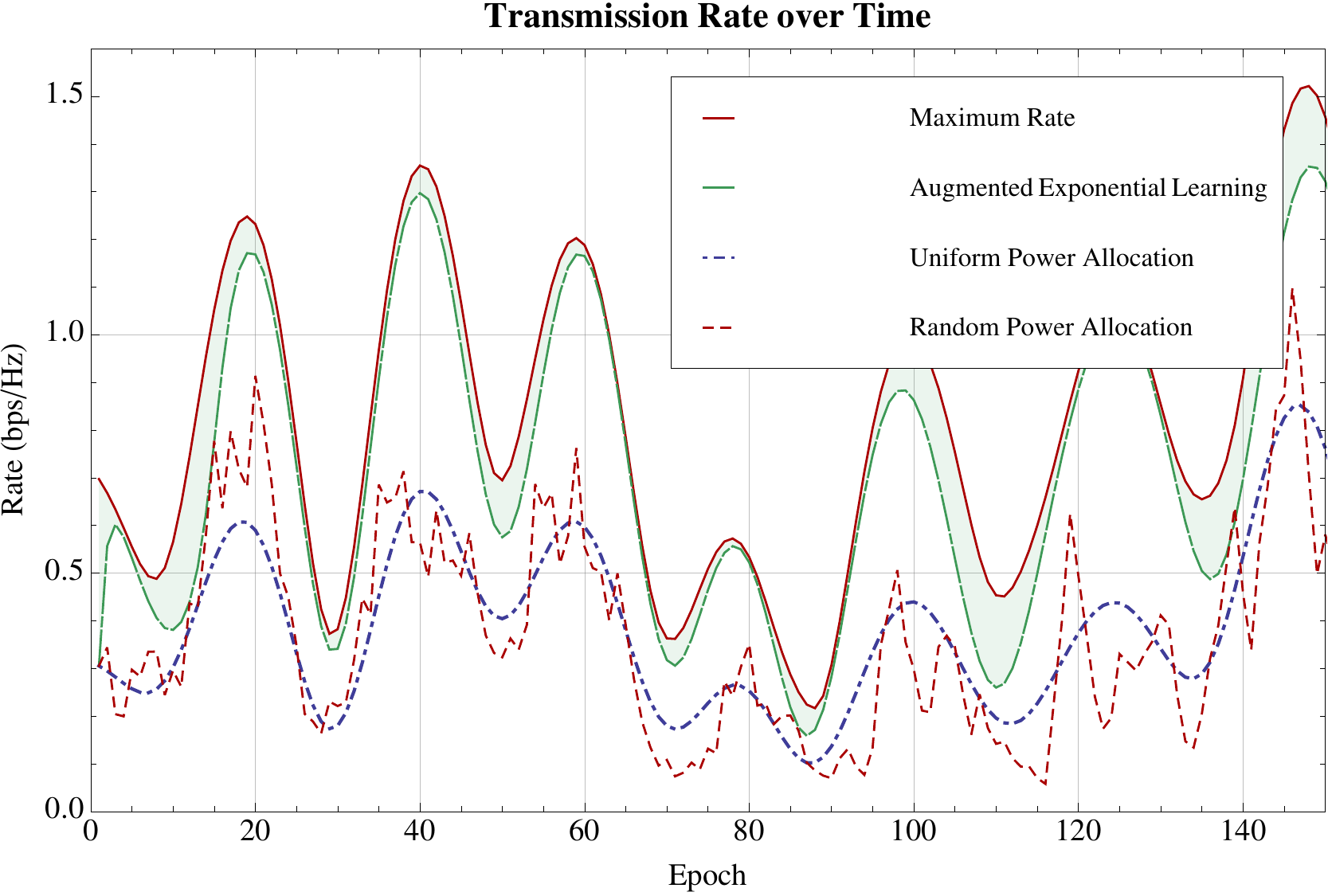}}
\hfill
\subfigure
[Performance of \ac{AXL} with average user velocity $v=15~\mathrm{km/s}$.]
{\label{fig:tracking-15}
\includegraphics[width=.485\textwidth]{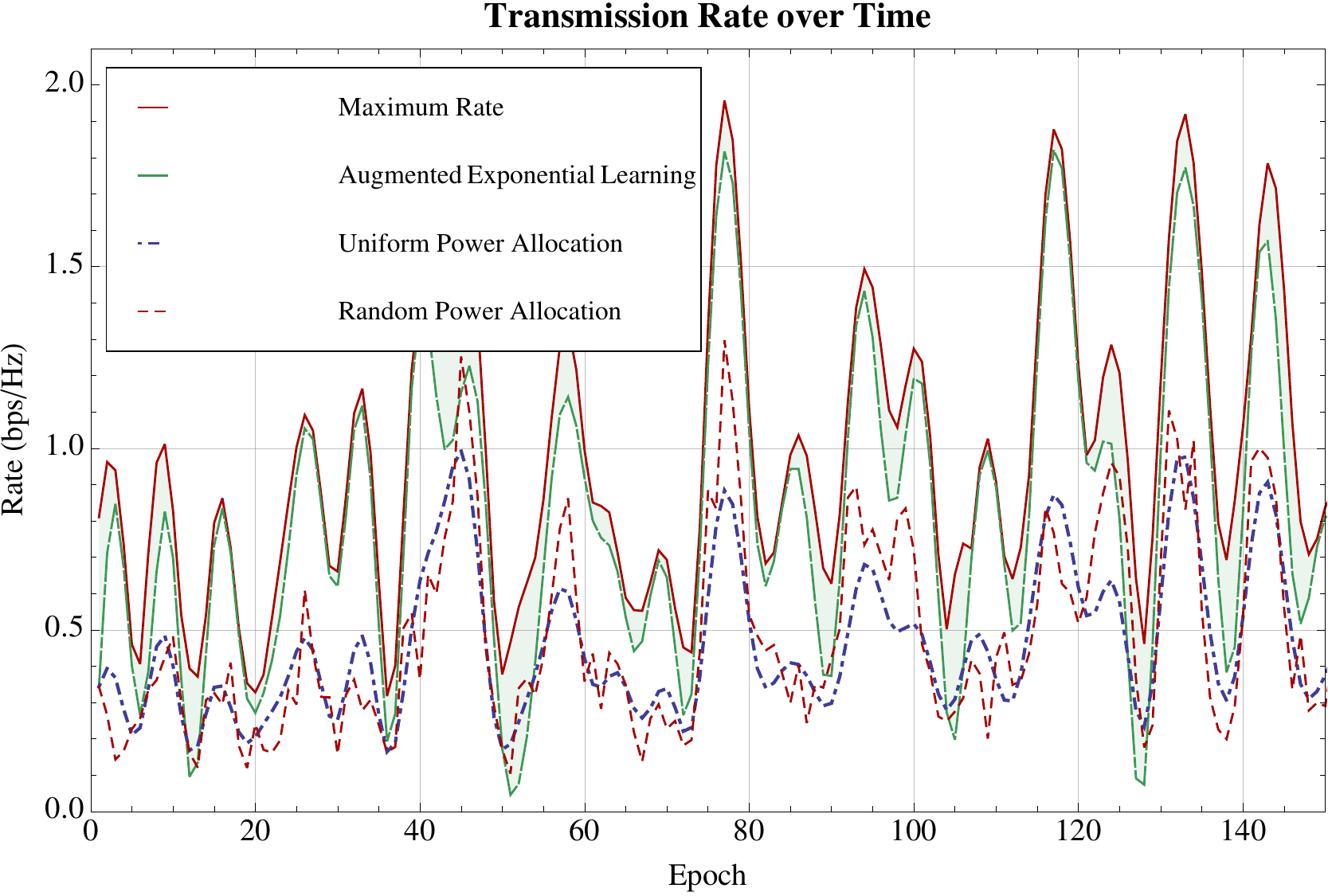}}
\caption{Data rates achieved by \ac{AXL} in a changing environment with different fading velocities:
the dynamic transmit policy induced by the \ac{AXL} algorithm allows users to track their maximum achievable transmission rate remarkably well even under rapidly changing channel conditions.}
\label{fig:tracking}
\end{figure*}

The negative-regret results of Fig. \ref{fig:regret} also suggest that the transmission rate achieved by a given \ac{SU} is close to the user's (evolving) maximum possible rate given the transmit profiles of every other user.
To test this hypothesis, we plotted in Fig.~\ref{fig:tracking} the achieved data rate of a \ac{SU} employing the \ac{AXL} algorithm along with the user's maximum achievable data rate and the rates achieved by the uniform policy and the randomized policy \eqref{eq:random};
to test different fading conditions, we simulated average user velocities of $v = 5~\mathrm{m/s}$ and $v = 15~\mathrm{m/s}$ (Figs. \ref{fig:tracking-5} and \ref{fig:tracking-15} respectively).
We see there that \ac{AXL} adapts to the changing channel conditions and tracks the user's maximum achievable rate remarkably well,
in stark contrast to the uniform and randomized transmit policies.%
\footnote{If the user's velocity becomes exceedingly high, the quality of this tracking may deteriorate as a result of the channel's extreme variability;
even in this case however, \ac{AXL} is guaranteed to perform at least as well as the best fixed transmit profile in hindsight.}

\begin{figure*}[htbp]
\centering
\subfigure
[Learning with perfect \ac{CSI}.]
{\label{fig:conv-det}
\includegraphics[width=.485\textwidth]{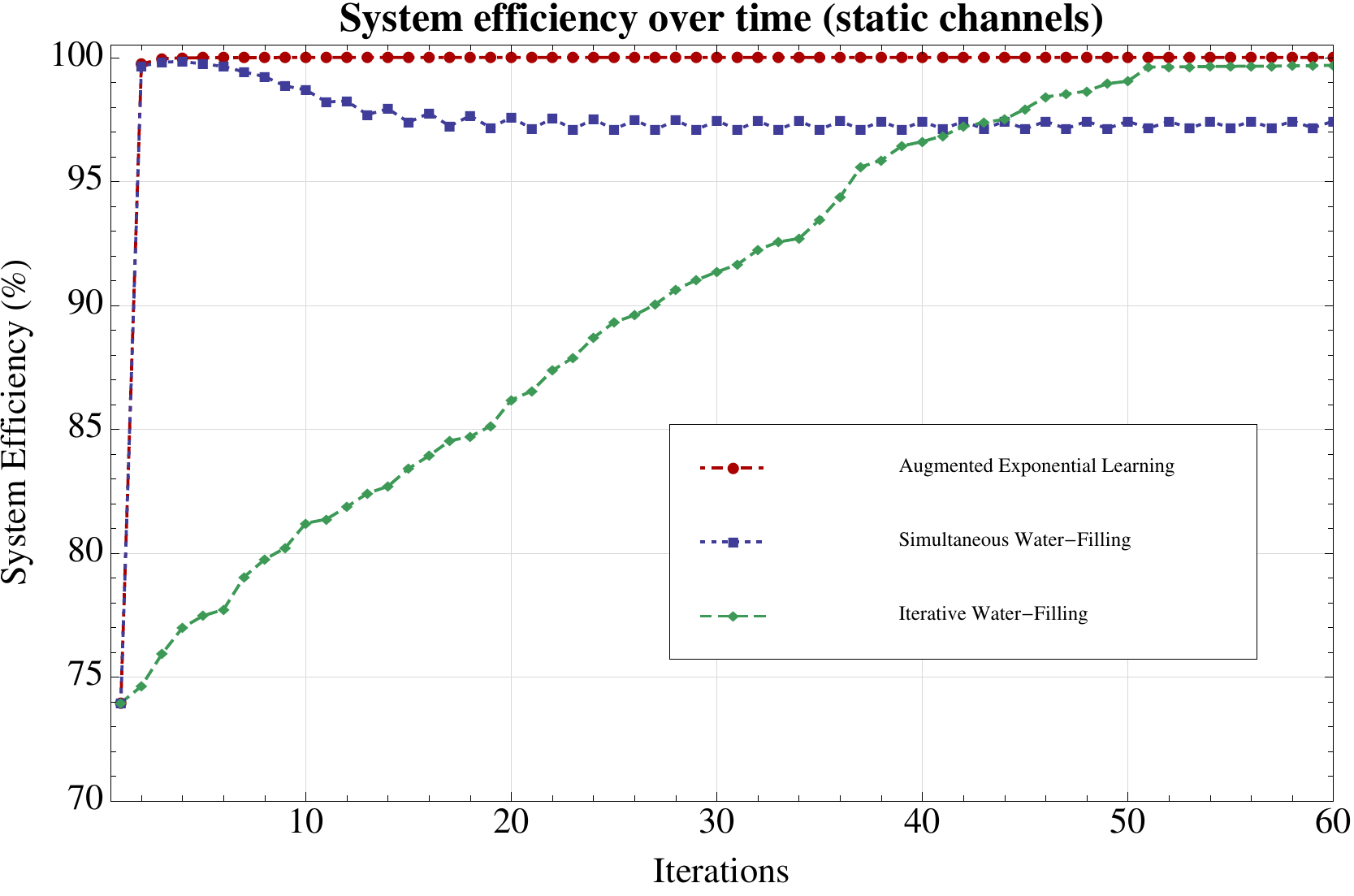}}
\hfill
\subfigure
[Learning with an average relative error level of 50\%.]
{\label{fig:conv-stoch}
\includegraphics[width=.485\textwidth]{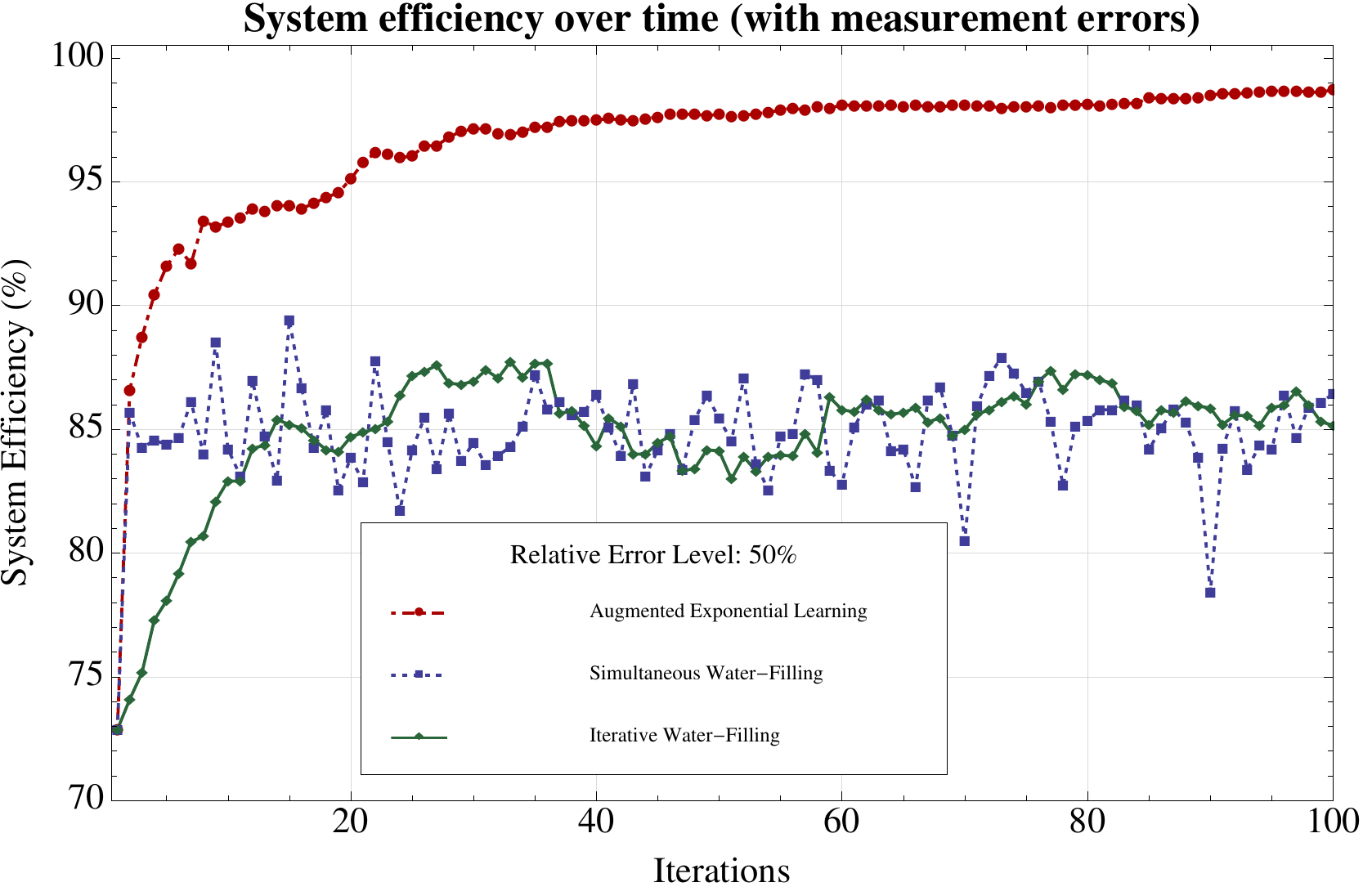}}
\caption{Convergence and robustness of \ac{AXL} with imperfect \ac{CSI} in a \ac{MIMO} \ac{MAC} system with $10$ \acp{PU} and $25$ \acp{SU}:
in contrast to water-filling methods, \ac{AXL} attains the channel's sum capacity even in the presence of very high measurement errors.}
\label{fig:convergence}
\end{figure*}

Finally, to assess the performance of the \ac{AXL} algorithm with respect to the users' sum rate under \acf{SIC} and the robustness of \ac{AXL} under imperfect \ac{CSI}, we simulated in Fig.~\ref{fig:convergence} a static multi-user \ac{MIMO} \acl{MAC} consisting of a wireless base receiver with $5$ antennas, $10$ \acp{PU} and $40$ \acp{SU} (each with a random number of transmit antennas picked uniformly between $2$ and $6$).
Each user's channel matrix $\bH_{k}^{qr} \equiv \bH_{k}^{q}$ was drawn from a complex Gaussian distribution at the outset of the transmission (but remained static once picked), and we then ran the \ac{AXL} algorithm with $\temp=1$.
The algorithm's performance over time was then assessed by plotting the \emph{efficiency ratio}
\begin{equation}
\label{eq:efficiency}
\eff(t) = \frac{\Psi(t) - \Psi_{\min}}{\Psi_{\max} - \Psi_{\min}},
\end{equation}
where $\Psi(t)$ denotes the users' sum rate at the $t$-th iteration of the algorithm, and $\Psi_{\max}$ (resp. $\Psi_{\min}$) is the maximum (resp. minimum) value of $\Psi$ over the set of feasible transmit profiles.%
\footnote{The reason for using this ratio was to eliminate scaling artifacts arising e.g. from the sum rate taking values in a narrow band close to its maximum value.}
For comparison purposes, we also plotted the efficiency ratio achieved by water-filling methods \textendash\ namely \ac{IWF} and \ac{SWF} \cite{SPB06}.
Remarkably, when the users have perfect \ac{CSI}, the \ac{AXL} policy achieves the system's maximum sum rate within $3$\textendash$4$ iterations;
by contrast, \ac{SWF} fails to converge altogether while the convergence time of \ac{IWF} scales linearly with the number of \acp{SU} (Fig.~\ref{fig:conv-det}).
On the other hand, in the presence of imperfect \ac{CSI} (modeled as zero-mean \acs{iid} Gaussian pertrubations to the gradient matrices $\gradpay_{k}$ with relative magnitude of $50\%$), \ac{AXL} still achieves the system's sum capacity (albeit at a slower rate) whereas water-filling methods offer no significant advantage over the user's initial transmit profile (cf. Fig.~\ref{fig:conv-stoch}).

\section{Conclusions}
\label{sec:conclusions}

In this paper, we introduced an adaptive transmit policy for \ac{MIMO}-\ac{OFDM} \acl{CR} systems that evolve dynamically over time as a function of changing user and environmental conditions.
Drawing on the method of matrix exponential learning \cite{MBM12} and online mirror descent \cite{SS11,KSST12}, we derived an \acf{AXL} scheme which leads to no regret:
for every \ac{SU}, the proposed transmit policy performs asymptotically as well as the best fixed transmit profile over the entire transmission horizon, and irrespective of how the system evolves over time.
In fact, this learning scheme is closely aligned to the direction of change of the users' data rate function, so the system's \acp{SU} are able to track their individual optimum transmit profile even under rapidly changing conditions.
Importantly, the implementation of the proposed algorithm requires only local \ac{CSI};
moreover, the algorithm retaints its no-regret properties even in the case of \emph{imperfect} \ac{CSI} (with arbitrarily large measurement errors) and significantly outperforms classical water-filling algorithms (where the use of perfect \ac{CSI} is critical).

To a large extent, our dynamic transmit policy owes its no-regret properties to an associated entropy function (for instance, the von Neumann quantum entropy for the problem's signal covariance component).
As a result, by choosing a proper entropy-like kernel (e.g. as in \cite{ABB04}), we can examine significantly more general situations, including for example pricing and/or energy-awareness constraints.

Finally, we should mention here that when the environment undergoes rapid changes, there are other regret notions which are more suited to adaptability (such as the adaptive regret measure of \cite{HazSes09}).
Studying the performance of \acl{AXL} with respect to different regret valuations lies beyond the scope of the current paper, but we intend to explore this direction in future work.

\appendix[Technical Proofs]

Our proof approach relies on a technique introduced by Sorin \cite{Sor09} and recently extended by J.~Kwon and one of the authors to more general \acl{OMD} methods \cite{KM14}.
First, we will establish the no-regret property of \acl{AXL} in continuous time;
subsequently, we derive the corresponding discrete-time result by estimating the difference between the continuous- and discrete-time processes.

\subsection{Online Power Allocation: the Case $P_{k}\geq P$.}
\label{app:PA}

To begin with, note that the exponential mapping of \eqref{eq:XLPA} may be characterized as the solution of the convex program:
\begin{equation}
\label{eq:Legendre-Gibbs}
\begin{aligned}
\text{maximize}
	\quad
	&\braket{\by}{\bq} - h(\bq),
	\\
\text{subject to}
	\quad
	&
	q_{k}\geq0,\; \sum_{k} q_{k} = 1,
\end{aligned}
\end{equation}
where $\braket{\by}{\bq}$ denotes the bilinear pairing $\braket{\by}{\bq} = \sum_{k} q_{k} y_{k}$ and $h(\bq) = \sum_{k} q_{k} \log q_{k}$ denotes the Gibbs\textendash Shannon entropy on the simplex $\simplex \equiv \simplex(\chan)$ spanned by $\chan$.
More precisely, we have the following classical result \cite[Chapter 25]{Roc70}:
\begin{lemma}
\label{lem:Gibbs}
For every $\by\in\R^{K}$, the problem \eqref{eq:Legendre-Gibbs} admits the unique solution $\gibbs(\by)$ with $\gibbs_{k}(\by) = e^{y_{k}} \big/ \sum_{\ell} e^{y_{\ell}}$.
\end{lemma}

Consider now the following continuous-time variant of \eqref{eq:XLPA} for $t\geq0$:
\begin{equation}
\label{eq:XLPA-cont}
\begin{aligned}
\dot\score_{k}
	&= \frac{\pd\Phi}{\pd q_{k}},
	\\
\bq(t)
	&= \gibbs\left(\vartemp(t) \vecscore(t)\right),
\end{aligned}
\end{equation}
where $\vartemp(t) = \min\{\temp,\temp t^{-1/2}\}$;
moreover, define the cumulative continuous-time regret with respect to some fixed $\testq\in\simplex$ as
\begin{equation}
\label{eq:reg-cont}
\reg_{T}^{c}(\testq)
	= \int_{0}^{T} \big[\rate(\testq;t) - \rate(\bq(t);t)\big] \dd t,
\end{equation}
where $\rate(\cdot;t)$, is a piecewise continuous stream of rate functions and the index $c$ in $\reg_{T}^{c}$ indicates that we are working in continuous time.
We then have:

\begin{proposition}
\label{prop:XLPA-cont}
The cumulative regret generated by the learning scheme \eqref{eq:XLPA-cont} satisfies
\(
\reg_{T}^{c}(\testq)
	\leq \temp^{-1} \log K \cdot \sqrt{T}
\)
for all $\testq\in\simplex$.
\end{proposition}

\begin{IEEEproof}
Let $h^{\ast}(\by)$ denote the convex conjugate of $h$, i.e.
\(
h^{\ast}(\by)
	= \max_{\bq\in\simplex} \{\braket{\by}{\bq} - h(\bq)\}
	= \braket{\vecscore}{\gibbs(\vecscore)} - h(\gibbs(\vecscore)).
\)
Moreover, set $\vartemp(t) = \min\{\temp t^{-1/2},\temp\}$ and let $\bq(t)$ be defined as in \eqref{eq:XLPA-cont} with $\vecpay(t) = \dot\vecscore(t) = \grad_{\bq(t)} \rate(\bq(t);t)$.
By Lemma \ref{lem:Gibbs}, we will have $h^{\ast}(\vartemp\vecscore) = \log \sum_{\ell} e^{\vartemp \score_{\ell}}$ and hence:
\begin{equation}
\label{eq:regPA-0}
\frac{d}{dt} h^{\ast}(\vartemp\vecscore)
	= \sum_{k\in\chan} \left.\frac{\pd h^{\ast}}{\pd \score_{k}}\right\vert_{\vartemp\vecscore}
	\left(\dot \vartemp \score_{k} + \vartemp \dot\score_{k}\right)
	= \dot\vartemp \braket{\vecscore}{\bq}
	+ \vartemp \braket{\vecpay}{\bq},
\end{equation}
where we used \eqref{eq:XLPA-cont} and the fact that $\grad_{\by} h^{\ast}(\by) = \gibbs(\by)$.
By isolating $\braket{\vecpay}{\bq}$ and integrating by parts, we then get:
\begin{flalign}
\label{eq:regPA-2}
\int_{0}^{T} \braket{\vecpay}{\bq} \dd t
	&= \frac{h^{\ast}(\vartemp(T) \vecscore(T))}{\vartemp(T)} - \frac{h^{\ast}(\vartemp(0) \vecscore(0))}{\vartemp(0)}
	+ \int_{0}^{T} \frac{\dot\vartemp}{\vartemp^{2}} h^{\ast}(\vartemp \vecscore) \dd t
	- \int_{0}^{T} \frac{\dot\vartemp}{\vartemp} \braket{\vecscore}{\bq} \dd t
	\notag\\
	\notag\\
	&= \frac{h^{\ast}(\vartemp(T) \vecscore(T))}{\vartemp(T)} - \frac{h^{\ast}(0)}{\vartemp(0)}
	- \int_{0}^{T} \frac{\dot\vartemp}{\vartemp^{2}} h(\gibbs(\vartemp \vecscore)) \dd t,
\end{flalign}
where the last step follows from the fact that $\bq = \gibbs(\vartemp \vecscore)$ and the defining relation $h^{\ast}(\vartemp \vecscore) = \braket{\vartemp\vecscore}{\gibbs(\vartemp\vecscore)} - h(\gibbs(\vartemp\vecscore))$.
Then, given that the minimum of $h$ over $\simplex$ is $-\log K$, we also have $h^{\ast}(0) = - h_{\min} = \log K$;
thus, with $\dot\vartemp \leq 0$, \eqref{eq:regPA-2} becomes:
\begin{flalign}
\label{eq:regPA-3}
\int_{0}^{T} \braket{\vecpay}{\bq} \dd t
	&\geq \frac{h^{\ast}(\vartemp(T) \vecscore(T))}{\vartemp(T)} - \frac{h^{\ast}(0)}{\vartemp(0)}
	+ h^{\ast}(0) \int_{0}^{T} \frac{\dot\vartemp}{\vartemp^{2}} \dd t
	\notag\\
	&\geq \frac{\braket{\vartemp(T)\vecscore(T)}{\testq} - h(\testq)}{\vartemp(T)} - \frac{\log K}{\vartemp(T)}
	\notag\\
	&\geq \braket{\vecscore(T)}{\testq} - \frac{\log K}{\temp} \sqrt{T},
\end{flalign}
where we used the fact that $h^{\ast}(\gamma\by) \geq \braket{\gamma\by}{\testq} - h(\testq)$ for all $\testq\in\simplex$ in the second line and that $h \leq 0$ in the last step.
With $\rate$ concave over $\simplex$, we will also have $\rate(\testq;t) - \rate(\bq(t);t) \leq \braket{\grad_{\bq(t)}\rate}{\testq-\bq(t)} = \braket{\vecpay(t)}{\testq - \bq(t)}$;
hence, by \eqref{eq:regPA-3}, we get:
\begin{equation}
\reg_{T}^{c}(\testq)
	\leq \int_{0}^{T} \braket{\vecpay}{\testq - \bq} \dd t
	\leq \frac{\log K}{\temp} \sqrt{T},
\end{equation}and our proof is complete.
\end{IEEEproof}


\subsection{Online Power Allocation: The General Case.}
\label{app:PA-mod}

If $P_{k}<P$ for some $k$,
we still obtain a no-regret power allocation policy if
we use the modified entropy function
\(
h(p) = \sum_{k} \left(p_{k} \log p_{k} + (P_{k} - p_{k}) \log (P_{k} - p_{k})\right),
\)
and define the modified Gibbs map:
\begin{equation}
\label{eq:choice-mod}
\gibbs_{0}(\by)
	= \argmax_{\bp\in\powset} \big\{\braket{\by}{\bp} - h_{0}(\bp)\big\}.
\end{equation}
Specifically, consider the following modified version of \eqref{eq:XLPA-cont}:
\begin{equation}
\label{eq:XLPA-mod-cont}
\begin{aligned}
\dot\score_{k}
	&= \frac{\pd\Phi}{\pd p_{k}},
	\\
\bp(t)
	&= \gibbs_{0}\left(\vartemp(t) \vecscore(t)\right),
\end{aligned}
\end{equation}
where $\rate(\cdot;t)$ is a continuous stream of rate functions of the form \eqref{eq:rate} and $\vartemp = \min\{\temp,\temp t^{-1/2}\}$.
We then have:
\begin{proposition}
\label{prop:XLPA-mod-cont}
The learning scheme \eqref{eq:XLPA-mod-cont} leads to no regret in continuous time:
\(
\reg_{T}^{c}(\testp)
	\leq \bigoh(\sqrt{T})
	\quad
	\text{for all $\testp\in\powset$.}
\)
\end{proposition}

\begin{IEEEproof}
As in the proof of Proposition \ref{prop:XLPA-cont}, let $h_{0}^{\ast}(\by) = \max_{\bp\in\powset}\{\braket{\by}{\bp} - h_{0}(\bp)\} = \braket{\by}{\gibbs_{0}(\by)} - h_{0}(\gibbs_{0}(\by))$ be the convex conjugate of $h_{0}(\bp)$.
Since the derivative of $h_{0}$ blows up to infinity at the boundary of $\powset$, the unique solution to the maximization problem defining $\gibbs_{0}$ lies at the interior of $\powset$.
The \ac{KKT} conditions thus give
\(
y_{k} - \frac{\pd h_{0}}{\pd p_{k}}
	= \lambda,
\)
where $\lambda$ is the Lagrange multiplier for the equality constraint $\sum_{\ell} p_{\ell} = P$.
We will then also have
\(
\frac{\pd h_{0}^{\ast}}{\pd y_{k}}
	= \gibbs_{0,k}(\by)
	+ \insum_{\ell=1}^{K} y_{\ell} \frac{\pd}{\pd y_{k}} \gibbs_{0,\ell} (\by)
	- \insum_{\ell=1}^{K} \frac{\pd h_{0}}{\pd p_{\ell}} \frac{\pd}{\pd y_{k}} \gibbs_{0,\ell}(\by)
	= \gibbs_{0,k}(\by),
\)
where, in the last step, we used
the fact that $\sum_{\ell=1}^{K} \gibbs_{0,\ell}(\by) = P$ (so $\sum_{\ell=1}^{K} \pd_{y_{k}} \gibbs_{0,\ell} = 0$ for all $k$).
Thus, letting $\vecpay(t) = \grad_{\bp} \rate(\bp;t)$ so that $\vecscore(t) = \int_{0}^{t} \vecpay(s) \dd s$ and $\bp(t) = \gibbs_{0}(\vartemp(t) \vecscore(t))$, we obtain the basic identity:
\begin{equation}
\label{eq:regPA-mod-0}
\frac{d}{dt} h_{0}^{\ast}(\vartemp\vecscore)
	= \sum_{k\in\chan} \left.\frac{\pd h_{0}^{\ast}}{\pd \score_{k}}\right\vert_{\vartemp\vecscore}
	\left(\dot \vartemp \score_{k} + \vartemp \dot\score_{k}\right)
	= \dot\vartemp \braket{\vecscore}{\bp}
	+ \vartemp \braket{\vecpay}{\bp},
\end{equation}
and the rest of the proof follows as in the case of Prop.~\ref{prop:XLPA-cont}.
\end{IEEEproof}

\subsection{Online Signal Covariance Optimization}
\label{app:COV}

For the \ac{MIMO} component \eqref{eq:OCOV} of \eqref{eq:ORM}
we will consider the continuous-time scheme:
\begin{equation}
\label{eq:XLCOV-cont}
\begin{aligned}
\dot\bscore_{k}
	&= \frac{\pd\rate}{\pd \bQ_{k}^{\ast}},
	\\
\bQ_{k}
	&= \frac{\exp(\vartemp \bscore_{k})}{\tr\left[\exp\left(\vartemp \bscore_{k}\right)\right]}.
\end{aligned}
\end{equation}
where, as before, $\vartemp = \min\{\temp,\temp t^{-1/2}\}$.
Then, with the user's regret defined as in \eqref{eq:reg-cont},
we get:

\begin{proposition}
\label{prop:XLCOV-cont}
The cumulative regret generated by the continuous-time learning scheme \eqref{eq:XLCOV-cont} satisfies
\(
\reg_{T}^{c}(\testQ)
	\leq \temp^{-1} \sqrt{T} \sum_{k=1}^{K} \log m_{k}
	\text{ for all $\txs \testQ\in\covset \equiv \prod_{k=1}^{K} \spectron_{k}$.}
\)
\end{proposition}

To prove Proposition \ref{prop:XLCOV-cont},
we first show that the matrix exponential of \eqref{eq:reg-XLCOV} solves the semidefinite problem:
\begin{equation}
\label{eq:Legendre-Neumann}
\begin{aligned}
\text{maximize}
	\quad
	&\tr\big[\bY \bQ\big] - h_{+}(\bQ),
	\\
\text{subject to}
	\quad
	&
	\bQ\mgeq0,\; \tr(\bQ) = 1,
\end{aligned}
\end{equation}
where $\bY$ is a Hermitian matrix and
\(
h_{+}(\bQ)
	= \tr\big[\bQ \log \bQ\big]
\)
is the \emph{von Neumann entropy}.
Indeed:
\begin{lemma}
\label{lem:Neumann}
For every Hermitian matrix $\bY\in\C^{m\times m}$, the problem \eqref{eq:Legendre-Gibbs} admits the unique solution $\bQ_{\bY} = \exp(\bY)\big/ \tr\big[\exp(\bY)\big]$.
Accordingly,
the convex conjugate $h_{+}^{\ast}$ of $h_{+}$ is:
\begin{equation}
h_{+}^{\ast}(\bY)
	= \max\nolimits_{\bQ\in\spectron} \left\{\tr\big[\bY \bQ\big] - h_{+}(\bQ)\right\}
	= \log \tr\big[\exp(\bY)\big].
\end{equation}
\end{lemma}

\begin{IEEEproof}
To begin with, let
\(
A(\bY,\bQ)
	= \tr\big[\bY \bQ] - h_{+}(\bQ)
\)
denote the objective of the problem \eqref{eq:Legendre-Neumann},
and let
\(
Z = \{\bA\in\C^{m\times m}: \bA^{\dag} = \bA, \tr(\bA) = 0\}
\)
be the space of tangent directions to $\spectron$.
Then, if $\{q_{j},\bu_{j}\}_{j=1}^{m}$ is an eigen-de\-com\-po\-si\-tion of $\bQ + t\bZ$ for $\bQ\in\intspectron$ and $\bZ\in Z$, we will have
\(
A(\bY,\bQ + t\bZ)
	= \tr[\bY \bQ] + \tr[\bY \bZ]\,t
	- \insum_{j} q_{j} \log q_{j}.
\)
Hence, the directional derivative of $A(\bY,\bQ)$ along $\bZ$ at $\bQ$ is
\(
\grad_{\bZ} A(\bY,\bQ)
	=\left.\frac{d}{dt}\right\vert_{t=0} A(\bY,\bQ+t\bZ)
	=\tr[\bY\bZ] - \insum_{k=1}^{K} \dot q_{k} \log q_{k}
\)
where we have used the fact that $\insum_{j} \dot q_{j} = 0$ (recall that $\insum_{j} q_{j} = \tr(\bQ + t\bZ) = 1$ for all $t$ such that $\bQ + t \bZ \in \intspectron$).
However, differentiating the defining relation $(\bQ + t\bZ) \bu_{j} = q_{j}\bu_{j}$ with respect to $t$ gives
\(
\bZ \bu_{j} + (\bQ + t\bZ) \dot \bu_{j}
	= \dot q_{j} \bu_{j} + q_{j} \dot\bu_{j},
\)
so, after multiplying from the left by $\bu_{j}^{\dag}$, we get
\(
\dot q_{j}
	= \bu_{j}^{\dag} \bZ \bu_{j}
	+ \bu_{j}^{\dag} (\bQ + t\bZ) \dot\bu_{j}
	- q_{j} \bu_{j}^{\dag} \dot\bu_{j}
	= \bu_{j}^{\dag} \bZ \bu_{j}.
\)
Summing over $j$ gives $\insum_{j}\dot q_{j}\log q_{j} = \insum_{j} \bu_{j}^{\dag} \bZ \bu_{j} \log q_{j} = \tr[\bZ \log \bQ]$;
then, by substituting in the previous expression for $\grad_{\bZ} A(\bY,\bQ)$, we finally obtain
\(
\grad_{\bZ} A(\bY,\bQ) = \tr[\bZ(\bY - \log \bQ)].
\)

By standard convex-analytic arguments, it follows that \eqref{eq:Legendre-Neumann} admits a unique solution $\bQ_{\bY}$ at the interior $\intspectron$ of $\spectron$ \cite[Chapter 26]{Roc70}.
Accordingly, by the \ac{KKT} conditions for \eqref{eq:Legendre-Neumann}, we have $\grad_{\bZ} A(\bY,\bQ_{\bY}) = 0$ for all tangent directions $\bZ$ to $\intspectron$ at $\bQ_{\bY}$, i.e.
\(
\tr[\bZ (\bY - \log \bQ_{\bY})] = 0
\)
for all Hermitian $\bZ \in \C^{m\times m}$ such that $\tr(\bZ) = 0$.
From this last condition, we immediately get $\bY - \log \bQ_{\bY} \propto \bI$, and with $\tr(\bQ_{\bY}) = 1$, we obtain $\bQ_{\bY} = \exp(\bY) / \tr[\exp(\bY)]$;
the expression for $h_{+}^{\ast}(\bY)$ then follows by substituting $\bQ_{\bY}$ in the definition of $A(\bY,\bQ)$.
\end{IEEEproof}

Armed with this characterization, we now get:

\begin{IEEEproof}[Proof of Proposition \ref{prop:XLCOV-cont}]
Let $h_{k}(\bQ_{k}) = \tr(\bQ_{k} \log \bQ_{k})$, $\bQ_{k}\in\spectron_{k}$, so $h_{k}^{\ast}(\bY_{k}) = \log \tr[\exp(\bY_{k})]$ by Lemma \ref{lem:Neumann};
moreover, let $\bQ = \diag(\bQ_{1},\dotsc,\bQ_{K})$ and set $h_{+}(\bQ) = \sum_{k} h_{k}(\bQ_{k}) = \tr\big[\bQ \log \bQ\big]$ for $\bQ\in\covset\equiv\prod_{k} \spectron_{k}$.
Then, if $\bY = \diag(\bY_{1},\dotsc,\bY_{K})$ with $\bY_{k}$ Hermitian, we will have
\(
h_{+}^{\ast}(\bY)
	= \max_{\bQ\in\covset} \left\{ \tr\big[\bY\bQ\big] - h(\bQ)\right\}
	= \insum_{k} h_{k}^{\ast}(\bY_{k})
	= \insum_{k} \log\tr\big[\exp(\bY_{k})\big].
\)
Accordingly, if we let $\bpay_{k}(t) = \pd_{\bQ_{k}^{\ast}}\rate(\bQ;t)$, we get:
\begin{flalign}
\label{eq:regCOV-1}
\frac{d}{dt} h_{+}^{\ast}(\vartemp\bscore)
	&= \insum_{k=1}^{K} \tr\big[\exp(\vartemp\bscore_{k})\big]^{-1} \frac{d}{dt} \tr\big[\exp(\vartemp\bscore_{k})\big]
	\notag\\
	&= \insum_{k=1}^{K} \tr\big[\exp(\vartemp\bscore_{k})\big]^{-1} \tr\big[\big(\dot\vartemp \bscore_{k} + \vartemp \dot\bscore_{k}\big) \exp(\bscore_{k})\big]
	\notag\\
	&= \dot\vartemp \tr\big[\bY \bQ\big] + \vartemp \tr\big[\bpay\bQ\big]
\end{flalign}
where we set $\bpay = \diag(\bpay_{1},\dotsc,\bpay_{K})$.
Following the same steps as in the proof of Proposition \ref{prop:XLPA-cont}, we then obtain:
\begin{equation}
\label{eq:regCOV-2}
\int_{0}^{T} \tr\big[\bpay \bQ\big] \dd t
	= \frac{h_{+}^{\ast}(\vartemp(T) \bscore(T))}{\vartemp(T)} - \frac{h_{+}^{\ast}(0)}{\vartemp(0)}
	- \int_{0}^{T} \frac{\dot\vartemp}{\vartemp^{2}} h_{+}(\bQ) \dd t,
\end{equation}
The minimum of $h_{+}$ over $\covset = \prod_{k}\spectron_{k}$ is just $-\sum_{k} \log m_{k}$, so we also have $h^{\ast}(0) = -\min_{\bQ\in\covset} h_{+}(\bQ) = \sum_{k} \log m_{k}$;
then, with $\dot\vartemp \leq 0$, \eqref{eq:regCOV-2} becomes:

\begin{flalign}
\label{eq:regCOV-3}
\int_{0}^{T} \tr\big[\bpay\bQ\big] \dd t
	&\geq \frac{h_{+}^{\ast}(\vartemp(T) \bscore(T))}{\vartemp(T)}
	- \frac{h^{\ast}(0)}{\vartemp(0)}
	+ h_{+}^{\ast}(0) \int_{0}^{T} \frac{\dot\vartemp}{\vartemp^{2}} \dd t
	\notag\\
	&\geq \frac{\tr\big[\vartemp(T)\bscore(T)\testQ\big] - h_{+}(\testQ)}{\vartemp(T)} - \frac{\sum_{k=1}^{K} \log m_{k}}{\vartemp(T)}
	\notag\\
	&\geq \tr\big[\bscore(T)\testQ\big] - \frac{\sum_{k=1}^{K} \log m_{k}}{\temp} \sqrt{T},
\end{flalign}
where we used the fact that $h_{+}^{\ast}(\gamma\bY) \geq \tr\big[\gamma\bY\testQ\big] - h_{+}(\testQ)$ for all $\testQ\in\covset$ in the second line and the fact that $h_{+} \leq 0$ in the last step.
Since $\rate$ is concave in $\bQ$ and $\bpay = \grad_{\bQ^{\ast}} \rate$, the rest of the proof follows in the same way as that of Proposition \ref{prop:XLPA-cont}.
\end{IEEEproof}

\subsection{The Full \ac{MIMO}\textendash\ac{OFDM} Problem}
\label{app:AXL}

Our final step in this continuous-time setting will be to establish the no-regret properties of the following continuous-time variant of the \ac{AXL} algorithm for $P_{k} \geq P$:
\begin{equation}
\label{eq:AXL-cont}
\begin{aligned}
\dot\score_{k}
	&= \frac{\pd\rate}{\pd q_{k}},
	&\hspace{0em}&
&\dot\bscore_{k}
	&= \frac{\pd\rate}{\pd \bQ_{k}^{\ast}},
	\\
q_{k}
	&= \frac{\exp(\vartemp \score_{k})}{\sum_{\ell=1}^{K}\exp(\vartemp \score_{\ell})},
	&\hspace{0em}&
&\bQ_{k}
	&= \frac{\exp(\vartemp \bscore_{k})}{\tr\left[\exp(\vartemp \bscore_{k})\right]},
\end{aligned}
\end{equation}
with $\vartemp = \min\{\temp,\temp t^{-1/2}\}$ as usual.
Without further ado, we have:

\begin{proposition}
\label{prop:AXL-cont}
If $P_{k}\geq P$ for all $k\in\chan$, then, for all $\testP\in\strat$, the cumulative regret generated by \eqref{eq:AXL-cont} will satisfy
\(
\reg_{T}^{c}(\testP)
	\leq \temp^{-1} \sqrt{T} \left(\log K + \sum_{k=1}^{K} \log m_{k}\right).
\)
\end{proposition}

\begin{IEEEproof}
Recall that any $\bP\in\strat$ may be decomposed as $\bP = \diag(p_{1}\bQ_{1},\dotsc,p_{K}\bQ_{K})$ with $\bp=(p_{1},\dotsc,p_{K})\in\powset$ and $\bQ = \diag(\bQ_{1},\dotsc,\bQ_{K}) \in \covset \equiv\prod_{k}\spectron_{k}$.
Then, using the normalized power allocation vector $\bq = \bp/P\in\simplex$ for convenience, let
\(
H(\bq,\bQ)
	= h(\bq) + h_{+}(\bQ)
	= \insum_{k=1}^{K} \big[q_{k} \log q_{k} + \tr(\bQ_{k} \log \bQ_{k})\big]
\)
and consider the associated Legendre\textendash Fenchel problem:
\begin{equation}
\label{eq:Legendre-agg}
\begin{aligned}
\textup{maximize}
	\quad
	&
	\braket{\by}{\bq} + \tr[\bY \bQ] - H(\bq,\bQ),
	\\
\textup{subject to}
	\quad
	&
	\bq\in\simplex,\;
	\bQ \in \inprod_{k}\spectron_{k}.
\end{aligned}
\end{equation}
Clearly, \eqref{eq:Legendre-agg} may be decomposed as a sum of \eqref{eq:Legendre-Gibbs} and \eqref{eq:Legendre-Neumann}, so each component of the solution of \eqref{eq:Legendre-agg} is given by Lemmas \ref{lem:Gibbs} and \ref{lem:Neumann} respectively;
likewise, the convex conjugate of $H$ will be $H^{\ast}(\by,\bY) = h^{\ast}(\by) + h_{+}^{\ast}(\bY)$, with $h^{\ast}$ and $h_{+}^{\ast}$ defined as before.
Our claim is then obtained by following the same steps as in the proofs of Propositions \ref{prop:XLPA-cont} and \ref{prop:XLCOV-cont}.
 \end{IEEEproof}

\subsection{The Descent to Discrete Time}
\label{app:discrete}

In this appendix, we
to derive the no-regret properties of the discrete-time policies \eqref{eq:XLPA}, \eqref{eq:XLCOV} and of the \ac{AXL} algorithm (Propositions \ref{prop:XLPA}, \ref{prop:XLCOV} and Theorem \ref{thm:AXL} respectively)
by means of a comparison technique introduced by Sorin \cite{Sor09} and developed further by J.~Kwon and one of the authors \cite{KM14}.
Specifically, we have:

\begin{lemma}
\label{lem:comparison}
Let $\cvset$ be a compact convex set in $\R^{N}$, let $\vecpay(t)$ be a sequence of payoff vectors in $\R^{N}$ with $\|\vecpay(t)\|\leq V$ in the uniform norm of $\R^{N}$ \textup($t=1,2\dotsc$\textup), and consider the sequence of play
\(
\bx(t+1)
	= \choice\left(\temp t^{-1/2} \insum_{s=1}^{t} \vecpay(s)\right)
\)
where $\choice\from\R^{N}\to\cvset$ is $C$-Lipschitz with respect to the $L^{1}$ norm on $\cvset$.
Moreover, letting $\vecpay^{c}(t) = \vecpay(\ceil{t})$ be a piecewise constant interpolation of $\vecpay(t)$ for $t\in[1,+\infty)$, consider the continuous-time process
\(
\bx^{c}(t)
	= \choice\left(\vartemp(t) \int_{0}^{t} \vecpay^{c}(s) \dd s\right)
\)
with $\vartemp(t) = \min\{\temp t^{-1/2}, \temp\}$, and assume that it guarantees the regret bound:
\begin{equation}
\label{eq:reg-bound-cont}
\int_{0}^{T} \braket{\vecpay^{c}(t)}{\testx - \bx^{c}(t)} \dd t
	\leq R(T) \sqrt{T}
	\quad
	\text{for all $\testx\in\covset$.}
\end{equation}
Then, for all $\testx\in\act$, the discrete-time sequence $\bx(t)$ guarantees
\begin{equation}
\label{eq:reg-bound-disc}
\insum_{t=1}^{T} \braket{\vecpay(t)}{\testx - \bx(t)}
	\leq \sqrt{T}\left(R(T) + 4 CV^{2} \temp\right).
\end{equation}
\end{lemma}

\begin{IEEEproof}
By assumption, if we set $\vecscore(t) = \int_{0}^{t} \vecpay^{c}(s) \dd s$, we have $\bx^{c}(t) = \choice(\vartemp(t)\vecscore(t)) = \bx(t+1)$ whenever $t$ is a positive integer.
Hence, for every integer $T\geq1$, we have
\(
\int_{0}^{T} \braket{\vecpay^{c}(t)}{\bx^{c}(t)} \dd t
	- \sum_{t=1}^{T} \braket{\vecpay(t)}{\bx(t)}
	= \int_{0}^{T} \braket{\vecpay^{c}(t)}{\bx^{c}(t)} \dd t
	- \int_{0}^{T} \braket{\vecpay(\ceil{t})}{\bx(\ceil{t})} \dd t
	= \int_{0}^{T} \braket{\vecpay^{c}(t)}{\bx^{c}(t) - \bx^{c}(\floor{t})} \dd t
\)
where we used the fact that $\bx^{c}(\floor{t}) = \bx(\ceil{t})$ in the second step.
On the other hand, H\"older's inequality gives
\(
\left\vert \braket{\vecpay^{c}(t)}{\bx^{c}(t) - \bx^{c}(\floor{t})} \right\vert
	\leq \|\vecpay^{c}(t)\|_{\infty} \cdot \|\bx^{c}(t) - \bx^{c}(\floor{t})\|_{1}
	\leq V\, \|\bx^{c}(t) - \bx^{c}(\floor{t})\|_{1}
	\leq V\, \|\choice(\vartemp(t) \vecscore(t)) - \choice(\vartemp(\floor{t}) \vecscore(\floor{t}))\|_{1}
	\leq CV\,\|\vartemp(t) \vecscore(t) - \vartemp(\floor{t}) \vecscore(\floor{t})\|_{\infty}.
\)
The last term may then be rewritten as:
\begin{flalign}
\label{eq:cont-disc-3}
\|\vartemp(t) \vecscore(t) - \vartemp(\floor{t}) \vecscore(\floor{t})\|_{\infty}
	&= \left\| \int_{\floor{t}}^{t} \frac{d}{ds} \left(\vartemp(s) \vecscore(s)\right) \dd s \right\|_{1}
	\\
	&\leq \int_{\floor{t}}^{t}
	\left\|
	\vartemp(s) \vecpay^{c}(s) + \dot\vartemp(s) \int_{0}^{s} \vecpay^{c}(w) \dd w\,
	\right\|_{\infty}
	\dd s
	\leq V\int_{\floor{t}}^{t} \left(\vartemp(s) - s \dot\vartemp(s)\right) \dd s.
\end{flalign}
Recalling that $\vartemp(t) = \min\{\temp,\temp t^{-1/2}\}$, this last integral is equal to $\temp t$ if $t\in[0,1]$ and $3 \temp\big(t^{1/2} - \floor{t}^{1/2}\big)$ otherwise.
Thus, combining the above inequalities, we obtain:
\begin{flalign}
\label{eq:cont-disc-4}
\int_{0}^{T} \negspace\braket{\vecpay^{c}(t)}{\bx^{c}(t) - \bx^{c}(\floor{t})} \dd t
	&\leq CV^{2} \negspace\int_{0}^{T} \negspace \int_{\floor{t}}^{t} \left(\vartemp(s) - s \dot\vartemp(s)\right) \dd s \dd t
	\\
	&\leq CV^{2} \temp
	\left(
	\frac{1}{2} + 3 \sum_{k=1}^{T-1} \int_{k}^{k+1} \frac{t-k}{\sqrt{t} + \sqrt{k}} \dd t
	\right)
	\leq 4 CV^{2} \temp \sqrt{T}.
\end{flalign}
Hence, by the definition of $\vecpay^{c}(t)$, we finally obtain
\begin{multline*}
\insum_{t=1}^{T} \braket{\vecpay(t)}{\testx - \bx(t)}
	= \int_{0}^{T} \braket{\vecpay^{c}(t)}{\testx - \bx^{c}(t)} \dd t
	+ \int_{0}^{T} \braket{\vecpay^{c}(t)}{\bx^{c}(t) - \bx^{c}(\floor{t})} \dd t
	\leq R(T) \sqrt{T} + 4 CV^{2} \temp \sqrt{T},
\end{multline*}
which completes our proof.
\end{IEEEproof}

With this comparison at hand, the analysis of the previous sections yields:

\begin{IEEEproof}[Proof of Proposition \ref{prop:XLPA}]
Note first that $\pay_{k} = \frac{\pd\rate}{\pd q_{k}} = P \tr\big[\gradpay_{k} \bQ_{k}\big]$, so the payoff vectors $\vecpay$ of \eqref{eq:pay-PA} are bounded in the uniform norm of $\R^{K}$ by $P\gradmax$ \textendash\ cf. \eqref{eq:paymax}.
Given that the Lipschitz constant of the exponential mapping $\gibbs(y)$ of \eqref{lem:Gibbs} is $C=1$ \cite{SS11}, the proposition follows by combining the continuous-time bound of Proposition \ref{prop:XLPA-cont} with Lemma \ref{lem:comparison}.
\end{IEEEproof}

\begin{IEEEproof}[Proof of Proposition \ref{prop:XLPA-mod}]
Note first that the modified Gibbs map of \eqref{eq:choice-mod} simply represents the power allocation policy of \eqref{eq:XLPA-mod}:
indeed, by the \ac{KKT} conditions for the maximization problem defining $\gibbs_{0}$, we will have:
\begin{equation}
\frac{p_{k}}{P_{k} - p_{k}} = e^{\lambda - y_{k}}
	\implies
p_{k} = P_{k} \frac{e^{y_{k}}}{e^{\lambda} + e^{y_{k}}},
\end{equation}
so, given that the power vector $\bp$ satisfies the total power constraint \eqref{eq:constraints-tot}, the Lagrange multiplier $\lambda$ must satisfy the condition $P = \sum_{k} p_{k} = \sum_{k} P_{k} (1 + e^{\lambda-y_{k}})^{-1}$.
Comparing this last equation with \eqref{eq:lambda}, we conclude that $p_{k}$ will be given by the power update step of \eqref{eq:XLPA-mod} with $\by$ replaced by $\vartemp\by$, so our claim follows by combining Proposition \ref{prop:XLPA-mod-cont} with Lemma \ref{lem:comparison}.
\end{IEEEproof}

\begin{IEEEproof}[Proof of Proposition \ref{prop:XLCOV}]
The matrix payoffs $\bpay_{k} = \frac{\pd\rate}{\pd\bQ_{k}^{\ast}} = p_{k} \gradpay_{k}$ satify $\|\bpay_{k}\| \leq P\gradmax$ by \eqref{eq:paymax}.
Moreover, the von Neumann entropy $h_{+}$ is $1$-strongly convex with respect to the $L^{1}$ norm, so the matrix exponential mapping $\bY\mapsto\bQ_{\bY} = \exp(\bY) \big/ \tr\big[\exp(\bY)\big]$ is $1$-Lipschitz \textendash\ see e.g. \cite{KSST12}.
Our claim then follows by combining the continuous-time bound of Proposition \ref{prop:XLCOV-cont} with Lemma \ref{lem:comparison}.
\end{IEEEproof}

\begin{IEEEproof}[Proof of Theorem \ref{thm:AXL}]
As in the proofs of Propositions \ref{prop:XLPA} and \ref{prop:XLCOV}, the map $(\by,\bY) \mapsto (\bq,\bQ)\in\simplex\times\prod_{k}\spectron_{k}$ of \eqref{eq:AXL-cont} is $1$-Lipschitz and the payoffs $(\vecpay,\bpay_{k})$ are bounded by $P\gradmax$ in the uniform norm of $\R^{K} \times \prod_{k} \C^{m_{k}\times m_{k}}$.
The theorem then follows by combining the continuous-time bound of Proposition \ref{prop:AXL-cont} with Lemma \ref{lem:comparison}.
\end{IEEEproof}

\subsection{Learning with Imperfect \ac{CSI}}
\label{app:stochastic}


\begin{IEEEproof}[Proof of Theorem \ref{thm:AXL-stoch}]
Let $\bP(t) = \diag\left(\bP_{1}(t),\dotsc,\bP_{k}(t)\right) \in\strat$ be the sequence of transmit profiles generated by the \ac{AXL} algorithm with noisy observations $\hat\gradpay = \gradpay + \bXi$.
Then, for every $\testP\in\strat$, we have:
\begin{multline}
\label{eq:reg-bound-stoch0}
\reg_{T}(\testP)
	\leq \insum_{t=1}^{T} \tr\left[\grad\rate(\bP(t)) \cdot \big(\testP - \bP(t)\big)\right]
	= \insum_{t=1}^{T} \tr\left[\hat\gradpay(t) \cdot \big(\testP - \bP(t)\big)\right]
	- \insum_{t=1}^{T} \tr\left[\bXi(t) \cdot \big(\testP - \bP(t)\big)\right],
\end{multline}
where the inequality follows from the concavity of $\rate$.
Since $\bP(t)$ is generated by the sequence of matrix payoffs $\hat\gradpay(t)$, the first term of this expression is simply the regret generated by $\bP(t)$ against $\hat\gradpay(t)$, so we have
\begin{equation}
\label{eq:reg-bound-stoch1}
\insum_{t=1}^{T} \tr\left[\hat\gradpay(t) \cdot \big(\testP - \bP(t)\big)\right]
	\leq R \sqrt{T}
\end{equation}
by Theorem \ref{thm:AXL} (or, more accurately, by combining \eqref{eq:regPA-3} and \eqref{eq:regCOV-3} with Lemma \ref{lem:comparison}).

As for the second term, it is easy to see that the process $V(t) = \tr\left[\bXi(t) \cdot \big(\bP(t) - \testP\big)\right]$ is a martingale difference:
indeed, since $\bP(t)$ is fully determined by $\hat\gradpay(1),\dotsc,\hat\gradpay(t-1)$, we get
\(
\ex[V(t) \vert \filter_{t-1}]
	= \ex\big[\tr\left[\bXi(t) \cdot \big(\bP(t) - \testP\big)\right] \vert \filter_{t-1}\big]
	=\tr\left[\ex\big[\bXi(t) \vert \filter_{t-1} \big] \cdot \big(\bP(t) - \testP\big)\right]
	= 0.
\)
Moreover, with $\|\bXi\| \leq \Sigma$, we will also have
\(
\vert V(t)\vert
	\leq \|\bXi(t)\| \cdot \|\testP - \bP(t)\|_{1}
	\leq \Sigma \cdot D,
\)
where $D = \max\{\|\testP - \bP\|_{1}:\testP,\bP\in\strat\}$ denotes the $L^{1}$-diameter of $\strat$.

The bound \eqref{eq:reg-stoch-avg} is thus obtained by taking the expectation of $\reg_{T}(\testP)$ and using the zero-mean property of $V$.
Similarly, the fact that $\bP(t)$ generates no regret almost surely (and not only in expectation) follows by noting that $T^{-1}\sum_{t=1}^{T} V(t) \to 0$ as a consequence of the strong law of large numbers for martingale differences \cite[Theorem 2.18]{HH80}.
Finally, for the large deviations bounds \eqref{eq:reg-stoch-largedev}, \eqref{eq:reg-bound-stoch0} yields:
\begin{equation}
\label{eq:reg-bound-largedev0}
\prob\left( \frac{1}{T}\reg_{T}(\testP) \geq \frac{R}{\sqrt{T}} + z\right)
	\leq \prob\left(\,\insum_{t=1}^{T} \vert V(t)\vert \geq T z \right).
\end{equation}
However, with $\|\bXi\| \leq \Sigma$, Azuma's inequality \cite{Azu67} yields
\(
\prob\left(\,\insum_{t=1}^{T} V(t) \geq Tz \right)
	\leq \exp\left(-\frac{T^{2}z^{2}}{2 \sum_{t=1}^{T} \ess\sup\vert V(t)\vert^{2}}\right)
	\leq \exp\left(- \frac{T z^{2}}{2 \Sigma^{2} D^{2}}\right),
\)
and our claim follows.
\end{IEEEproof}


\balance
\bibliographystyle{IEEEtran}
\footnotesize
\bibliography{IEEEabrv,Bibliography}

\end{document}